\documentclass[aps,prb,twocolumn,showpacs,superscriptaddress,floatfix]{revtex4}

\usepackage{bm}
\usepackage{color}
\usepackage{amssymb}
\usepackage{amsmath}
\usepackage{graphicx}
\usepackage{amsmath}
\usepackage{dcolumn}
\usepackage{bm}
\usepackage{bbm}
\usepackage[english]{babel}

\newcommand{\beq}{\begin{eqnarray}}
\newcommand{\eeq}{\end{eqnarray}}
\newcommand{\beqa}{\begin{equation}}
\newcommand{\eeqa}{\end{equation}}

\begin{document}

\title{Modeling Klein tunneling and caustics of electron waves in graphene}
\author{R. Logemann}
\affiliation{Radboud University of Nijmegen, Institute for Molecules and Materials,
Heijendaalseweg 135, 6525 AJ Nijmegen, The Netherlands}
\author{K. J. A. Reijnders}
\affiliation{Radboud University of Nijmegen, Institute for Molecules and Materials,
Heijendaalseweg 135, 6525 AJ Nijmegen, The Netherlands}
\author{T. Tudorovskiy}
\affiliation{Radboud University of Nijmegen, Institute for Molecules and Materials,
Heijendaalseweg 135, 6525 AJ Nijmegen, The Netherlands}
\author{M. I. Katsnelson}
\affiliation{Radboud University of Nijmegen, Institute for Molecules and Materials,
Heijendaalseweg 135, 6525 AJ Nijmegen, The Netherlands}
\author{Shengjun Yuan}
\email{s.yuan@science.ru.nl}
\affiliation{Radboud University of Nijmegen, Institute for Molecules and Materials,
Heijendaalseweg 135, 6525 AJ Nijmegen, The Netherlands}
\date{\today}

\begin{abstract}
We employ the tight-binding propagation method to study Klein tunneling and quantum interference in large graphene systems. With this efficient numerical scheme, we model the propagation of a wave packet through a potential barrier and determine the tunneling probability for different incidence angles. We consider both sharp and smooth potential barriers in \emph{n-p-n} and \emph{n-n'} junctions and find good agreement with analytical and semiclassical predictions. When we go outside the Dirac regime, we observe that sharp $n$-$p$ junctions no longer show Klein tunneling because of intervalley scattering. However, this effect can be suppressed by considering a smooth potential. Klein tunneling holds for potentials changing on the scale much larger than the interatomic distance. When the energies of both the electrons and holes are above the Van Hove singularity, we observe total reflection for both sharp and smooth potential barriers. Furthermore, we consider caustic formation by a two-dimensional Gaussian 
potential. For sufficiently broad potentials we find a good agreement between the simulated wave density and the classical electron trajectories.
\end{abstract}

\maketitle

\section{Introduction}

Graphene, a single layer of carbon atoms arranged in a honeycomb lattice, has attracted great interest because of its special electronic properties. These special properties result from the fact that its charge carriers satisfy the massless Dirac equation.\cite{Graphene-review,Beenakker2008,P10,DasSarma2011,KatsnelsonBook}. One of these unique properties is the unusual tunneling of electrons through energy barriers, so-called Klein tunneling\cite{Klein29,katsnelson-nature-2006,Cheianov06}: For an electron that is normally incident on a potential barrier, there will always be total transmission of the electron, independent of its kinetic energy and of the height and width of the potential barrier. This is in contrast to usual quantum tunneling, where the tunneling probability decays exponentially as a function of the barrier height and width. 
The origin of Klein tunneling is the existence of an additional degree of freedom (pseudospin) which is conserved across the barrier interface.\cite{katsnelson-nature-2006,Allain11,Tudorovskiy13} Earlier, the absence of back scattering for massless Dirac fermions was considered in terms of the Berry phase, in the context of carbon nanotubes.\cite{Ando98}
Soon after its theoretical prediction, Klein tunneling in graphene was observed by several experimental groups.\cite{Stander2009,Young2009}
Recently, angular scattering by a graphene \emph{p-n} junction was also studied experimentally.~\cite{Sutar12}

In this paper, we study Klein tunneling and other scattering processes in graphene numerically using the tight-binding propagation method (TBPM),\cite{YRK10,WK10,YRRK11,Yuan2012} which has its origins in Ref.~\onlinecite{Hams00}. Given an initial wave packet, the method determines its time evolution on the graphene lattice by solving the time-dependent Schr\"odinger equation (TDSE) for the tight-binding Hamiltonian. Because of its efficient implementation, the computational time and memory required scale only linearly with system size, allowing the study of large systems.

In Ref.~\onlinecite{Pereira2010}, Klein tunneling in graphene was studied numerically for both a single barrier and for multiple barriers. By solving the time-dependent Schr\"odinger equation for the Dirac Hamiltonian using the split-operator method, the authors studied the propagation of a Gaussian wave packet. Because this wave packet was much smaller than the size of the graphene sample, the authors could directly obtain the reflection and transmission angle of the wave packet. However, since a Gaussian wave packet contains components with different wave vectors, one cannot extract the reflection and transmission as a function of the wave vector from such a calculation. Since these are the quantities that are usually determined in a theoretical analysis,\cite{katsnelson-nature-2006,Cheianov06,Shytov08,Tudorovskiy13,Reijnders14} it is difficult to compare the computational results to theoretical predictions. Other numerical studies of scattering of Gaussian wave packets were performed in Refs.~\onlinecite{Rakhimov11,Palpacelli12}.

In our approach, to prepare the initial wave packet, we take a plane (sinusoidal) wave with a given wave vector and cut from it only a finite part, with a total length $L$ of about 10-20 wavelengths on average. In the rest of the text, we will call such an object a ``plane wave packet''.
Although it necessarily contains additional wave vectors with a distribution width of the order of $2\pi/L$, their amplitude is small and the wave packet is a good approximation to a plane wave. 
Because the TBPM permits the study of large systems, we can use it to study the propagation of this large wave packet. We explain our method in more detail in Section~\ref{sec:method}.

We apply our numerical scheme to two different cases. In Section~\ref{sec:KT}, we first study angular scattering for one-dimensional \emph{n-p-n} and \emph{n-n'} junctions in the Dirac regime. For different angles of incidence, we extract the transmission and compare it with theoretical results. For the sharp junction, the latter can be obtained by exact wave matching at the barrier interface.~\cite{katsnelson-nature-2006,Allain11} For smooth potentials, we use semi-analytical formulas that were recently derived using the semiclassical approximation.\cite{Tudorovskiy13,Reijnders14} However, our numerical scheme is not limited to the Dirac regime, and we also consider the transmission through both sharp and smooth $n$-$p$ junctions for various energies outside this regime. In particular, we investigate whether the exact 100\% transmission for a normally incident electron persists or is no longer present. The latter happens when the next-nearest-neighbor hopping $t'$ is introduced.\cite{Kretinin13} 
Furthermore, we also pay special attention to what happens to the transmission near the Van Hove singularity. It has been shown that the character of the quantum Hall effect changes abruptly when passing this point because of the change in the Chern number.\cite{Hatsugai06} Therefore, there may also be a change in the tunneling behavior. 

The second application of our method is given in Section~\ref{sec:caust}, where we discuss scattering by a two-dimensional potential whose maximum is lower than the energy of the wave packet. The main effect of this potential is that the (classical) electron trajectories are bent, which leads to focusing. The envelope of the trajectories is known as a caustic, and corresponds to a region of increased intensity. In the literature, focusing of electrons in graphene has mainly been discussed in the context of \emph{n-p} or \emph{n-p-n} junctions.\cite{Cheianov07,Cserti07,Wu14} Focusing by such junctions is analogous to focusing by a lens with a negative refractive index, which opens up the possibility of realizing the electron analog of a so-called Veselago lens.\cite{Cheianov07}
In Ref.~\onlinecite{Wu14}, the authors also studied scattering of electrons above a sharp circularly symmetric potential. 
Furthermore, in Ref.~\onlinecite{Guimaraes11}, a method was proposed to focus spin currents in graphene instead of the electronic current.
However, we will only be concerned with above-barrier scattering of electrons, and we will compare the intensity in the area behind the potential with the classical electron trajectories and the associated caustic.

In Section~\ref{sec:conclusion}, we give an overview of the main results and discuss possibilities for future work.

\section{Method and Model} \label{sec:method}

In this section, we discuss the details of the model and the computational scheme.

\subsection{Tight-binding Hamiltonian}

For single-layer graphene, the tight-binding (TB) Hamiltonian in the single $\pi$-band model (which is sufficient to describe the electronic structure of graphene in a broad interval, plus minus several electronvolts around the Dirac point\cite{KatsnelsonBook}) is given by
\begin{equation}
H=-\sum_{<i,j>}t_{ij}c_{i}^{\dagger }c_{j}+\sum_{i}v_{i}c_{i}^{\dagger }c_{i} ,
\label{Eq:SLG}
\end{equation}%
where $t_{ij}$ is the nearest neighbor hopping parameter between sites $i$
and $j$, $c_{i}^{\dagger }$ and $c_{i}$ are the creation and annihilation
operators at site $i$ and $v_{i}$ is the on-site potential.
For pristine graphene, the nearest neighbor hopping is uniform, so that $t_{ij}\equiv
t=3.0$~eV, and the on-site potential is zero ($v_{i}=0$).
For an infinite graphene system, the TB Hamiltonian~(\ref{Eq:SLG}) is diagonalized by the Bloch eigenstates
\begin{equation}
|\mathbf{k}\rangle =\sum_{i}a_{i}c_{i}^{\dagger }|0\rangle  \label{Eq:WaveK}
\end{equation}%
where%
\begin{equation}
a_{i}=\left\{
\begin{array}{ll}
\frac{e^{i\mathbf{k}\cdot \mathbf{r}_{i}}}{\sqrt{2}}\frac{\lambda f(\mathbf{k%
})}{|f(\mathbf{k})|}, & i\in \text{Sublattice A} \\
\frac{e^{i\mathbf{k}\cdot \mathbf{r}_{i}}}{\sqrt{2}}\text{ \ \ }, & i\in
\text{Sublattice B}%
\end{array}%
\right. . \label{eq:coeffs_ai}
\end{equation}%
The function $f\left( \mathbf{k}\right) $ is defined as%
\begin{equation}
f\left( \mathbf{k}\right) =\exp \left( -i\mathbf{k}\cdot\boldsymbol{\delta}_{1}\right) +
\exp \left( -i\mathbf{k}\cdot\boldsymbol{\delta}_{2}\right) +
\exp \left( -i\mathbf{k}\cdot\boldsymbol{\delta}_{3}\right) , \label{eq:energy_f}
\end{equation}%
where $\boldsymbol{\delta}_{i}$ are vectors pointing to the three nearest neighbors
of an atom in the honeycomb lattice:
\begin{equation}
\boldsymbol{\delta}_{1} =\frac{a}{2}\left(\sqrt{3},1\right) , \; \boldsymbol{\delta}_{2} =\frac{a}{2}\left( -\sqrt{3},1\right) , \;
\boldsymbol{\delta}_{3} =-a\left( 0,1\right) ,
\end{equation}
with $a \approx 1.42$ $\text{\AA}$ the spacing between two carbon atoms.
The constant $\lambda$ takes the values $\pm1$, giving rise to two bands, which are referred to as the $\pi^{\ast}$ and $\pi$
bands.
The eigenenergy of the state $|\mathbf{k}\rangle $ equals
\begin{equation}
E\left( \mathbf{k}\right) =\lambda t\left\vert f\left( \mathbf{k}\right) \right\vert ,
\label{Eq:EnergyK}
\end{equation}%
where $\mathbf{k}$ is the wave vector with respect to the center of the Brillioun zone.

At the conical points
\begin{equation}
  \mathbf{K}=\left( \frac{4\pi}{3\sqrt{3} a}, 0 \right) \quad \text{and} \quad \mathbf{K'}=\left( -\frac{4\pi}{3\sqrt{3} a}, 0 \right) , \label{eq:KKpr}
\end{equation}
the energy $E\left( \mathbf{k}\right)$ vanishes and the two bands touch. In the neighborhood of these points the energy is linear in the wave vector $|\mathbf{k}|$, \mbox{$E(k)=\hbar v_F |\mathbf{k}|$}, where $v_F=3 t a/2 \approx c/300$ is called the Fermi velocity. For energies below 1~eV, the Hamiltonian can be approximated by the massless Dirac Hamiltonian,
\begin{equation}
  \hat{H} = v_F \boldsymbol{\sigma} \cdot \hat{\mathbf{p}} + U(x,y) , \label{eq:Dirac}
\end{equation}
where $\boldsymbol{\sigma}=(\sigma_x,\sigma_y)$ is the vector of Pauli matrices and $\hat{\mathbf{p}}=(\hat{p}_x,\hat{p}_y)$ are the momentum operators \mbox{$\hat{p}_x=-i \hbar \partial /\partial x$}. The external potential $U(x,y)$ is zero for pristine graphene.

Note that in the remainder of this paper, we measure the wave vector $\mathbf{k}$ with respect to the $\mathbf{K}$-point.

\subsection{Preparation of the wave packet}

The wave function expressed in Eq.~(\ref{Eq:WaveK}) is a plane wave defined on an infinite lattice. Because numerical models cannot handle infinite systems in real space, we need to find an approximate way to introduce the wave vector $\mathbf{k}$ in the simulation. One way is to introduce a Gaussian wave packet as was done in Ref.~\onlinecite{Pereira2010}:%
\begin{equation}
\psi \left( x,y\right) =\frac{1}{\delta \sqrt{2\pi }}\exp \left[ -\frac{%
\left( x-x_{0}\right) ^{2}}{2\delta ^{2}}-\frac{\left( y-y_{0}\right) ^{2}}{%
2\delta ^{2}}+ik_{x}x\right] .
\end{equation}%
Using this finite-sized Gaussian wave packet, the reflection and transmission angles at an \emph{n-p} junction can be measured directly from the direction of the reflected and transmitted wave. On the other hand, as mentioned before, a small-sized Gaussian wave packet differs a lot from a plane wave that is used in theoretical studies of Klein tunneling.

\begin{figure}[tbp]
\centering
\includegraphics[width=1.0\columnwidth]{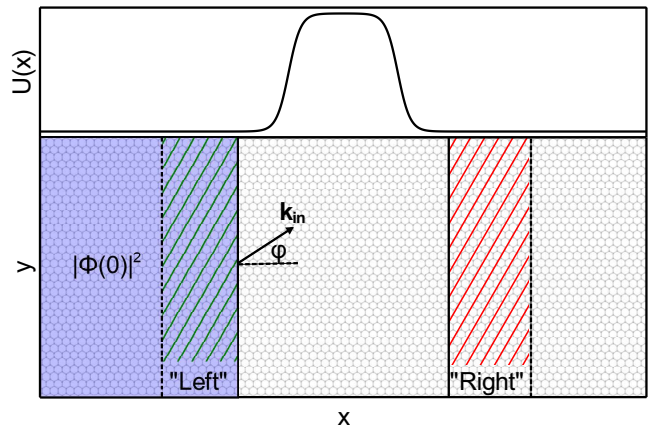}
\caption{Computational setup for the simulation of an \mbox{n-p-n} junction. The initial wave packet is localized on the left of the lattice
and is indicated in grey/blue. Its wave vector $\mathbf{k}_{in}$ makes an angle $\protect\varphi$ with the $x$-axis. The junction is located at the center of the lattice. To obtain the transmission and reflection, densities in the ``Left'' and ``Right'' regions are calculated as a function of time.}
\label{fig:simulation_setup_npn}
\end{figure}

In our numerical simulations, the initial wave packet is created exactly according to Eq.~(\ref{Eq:WaveK}). The setup of our numerical simulations is shown in Fig.~\ref{fig:simulation_setup_npn}. Since we would like to study the propagation of the wave, the initial wave packet is localized in one part (in our case on the left side) of the graphene sample, which means that the summation over $i$ in Eq.~(\ref{Eq:WaveK}) is restricted to this region only. 
The wave vector is chosen to have positive $k_{x}$, so that the wave will propagate from left to right. We use periodic boundary conditions in the $y$-direction and open boundary conditions in the $x$-direction. The periodic boundary conditions in the $y$-direction are necessary in order to prevent reflections from the boundaries.
Whenever the transversal wave vector $k_y$ is nonzero, the length $L_y$ of the sample in the $y$-direction is chosen in such a way that $L_y/\lambda_y=L_{y}k_{y}/2\pi$ is as close to an integer as possible in order to match the phases at the top and bottom edges.
Note that in the presence of periodic boundary conditions any mismatch of the phases at these edges would introduce extra interference terms during the wave propagation, which will affect the values of the transmission and reflection probabilities.

\subsection{Tight-binding propagation method}

The next step of our procedure is to calculate the propagation of the wave packet along the sample according to the time-dependent Schr\"{o}dinger equation (TDSE):
\begin{equation}
\left\vert \Phi \left( t\right) \right\rangle =e^{-i H t/\hbar}\left\vert \Phi
\left( 0\right) \right\rangle ,
\end{equation}%
For a general initial state $\left\vert \Phi \left( 0\right) \right\rangle $, the action of the time evolution operator $e^{-i H t/\hbar}$ for the TB Hamiltonian is calculated numerically by using the Chebyshev polynomial algorithm\cite{YRK10,WK10,Abramowitz65,Press07}. This so-called tight-binding propagation method (TBPM) is extremely efficient, because the TB Hamiltonian is a sparse matrix.\cite{Hams00}
Furthermore, it has the advantage that the CPU time and memory cost are both linearly dependent on the system size. For more details and examples of the numerical calculation of the time-evolution operator for graphene based systems we refer to Refs.~\onlinecite{YRK10,WK10,YRRK11,Yuan2012}. 
Using the TBPM, we find the spatial distribution of the wave packet density at each timestep. The simulation is stopped when the wave front reaches the right side of the sample.

To check the validity of our setup, we simulated the propagation of a wavepacket through a graphene sample without an external potential. 
\begin{figure}[tbp]
\centering
\mbox{
\includegraphics[width=1.0\columnwidth]{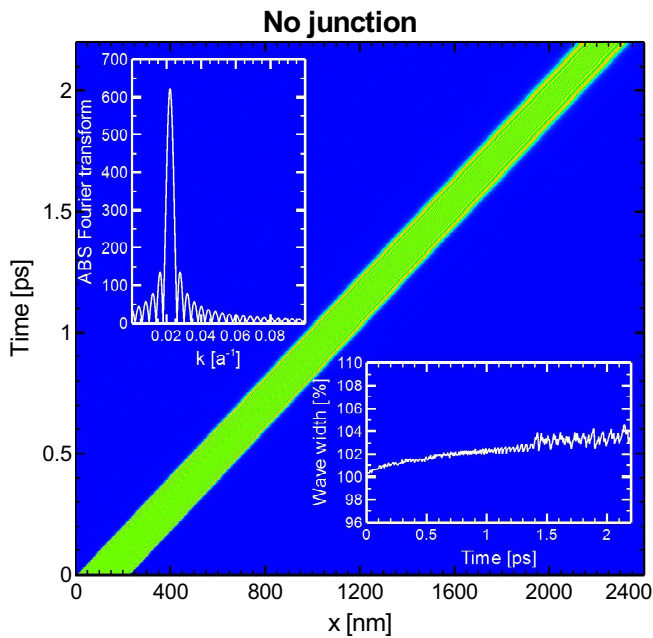}
}
\caption{The evolution of the density, integrated along the $y$-direction for every $x$ as a function of time. The wave packet, which has an energy of $0.1$~eV, propagates from left to right with only very little dispersion. The bottom right inset shows the width of the wave packet as a function of time relative to $t=0$. The top left inset shows the Fourier Transform of the initial wave packet. The full width at half maximum of the central peak is $0.005 \; a^{-1}$, which is approximately equal to $2 \pi/L$.}
\label{fig:no_juction}
\end{figure}
To monitor the time evolution of the wave packet in our numerical simulations, we integrate the density along the $y$-direction for every point $x$ at each time $t$. In Fig.~\ref{fig:no_juction}, where this integrated density is shown, we see that the width of the wave packet is approximately constant and only increases by approximately 4\%. Furthermore, the probability density remains homogeneous in the middle of the wave packet, which implies that the center of the wave packet is a good approximation to a plane wave with a certain wave vector $k_0$. At the edges, the probability density is less homogeneous, and we see the influence of the additional wave vectors that are introduced because of the finite width of the wave packet. In the top left inset of Fig.~\ref{fig:no_juction}, we show the Fourier transform of the initial wave packet. We see that it has a sharp peak around $k_0=0.022 \; a^{-1}$, with a full width at half maximum of $0.005 \; a^{-1}$, which approximately equals $2 \pi/L$. We remark that this wave packet is among the smallest that we have used in our simulations.

We note that near the Dirac point the dispersion is linear, and hence all wave vectors have the same phase velocity. Therefore, only $k$-vectors that correspond to energies outside of the Dirac regime contribute to the broadening of the wave packet. This implies that when our energy is in the Dirac regime, the wave packet propagates like a classical wave packet with only very little dispersion. This behavior is indeed seen in Fig.~\ref{fig:no_juction}.

For our second application, where we study focusing of electrons by a two-dimensional potential, the wave packet density is what we are interested in. For our first application, where we study angular scattering by one-dimensional \emph{n-p-n}, \emph{n-n'} and  \emph{n-p} junctions, we still have to extract the transmission from this data.

\subsection{Extracting the transmission}

We discuss two ways of extracting the transmission. The first method is mainly suitable for \emph{n-p-n} junctions, whereas the second method works for \emph{n-n'} and \emph{n-p} junctions. Note that in order for the reflection and transmission to be well-defined, we require that the potential is constant on the left and on the right of the potential barrier.

\subsubsection{n-p-n junction}  \label{sec:method_npn}

In this method, we start by choosing two small regions of the same width, on the left and on the right of the junction, in the region where the potential is constant, as indicated in Fig.~\ref{fig:simulation_setup_npn}.
The sum $\sum_{i}|\psi_i(t)|^2$ of the wave density over all sites in a certain region is denoted as the wave amplitude in that region.
The wave amplitude in the left region at the initial time is regarded as the amplitude $A_{in}$ of the incoming wave, and the wave amplitude in the right region is the time-dependent wave amplitude $A_{out}\left( t\right)$ of the transmitted wave. When the potential in the left and right region is the same, the transmission at time $t$ can be calculated as
\begin{equation}
  T\left( t\right) =\frac{A_{out}\left( t\right) }{A_{in}} . \label{eq:defTrans}
\end{equation}
It is important to note that because of the two barrier interfaces, there are internal reflections within the barrier and the total transmission can be represented as a sum of multiscattering processes. Therefore, the transmission $T$ increases over time, and one obtains the transmission as
\begin{equation}
T=\lim_{t\rightarrow \infty }T\left( t\right) .
\label{eq:transmission_infinite}
\end{equation}
However, one can only consider infinite times in Eq.~(\ref{eq:transmission_infinite}) if the system is \emph{infinitely} large in the $x$-direction. For a finite system, the wave packet will bounce from the right side of the sample, and one should measure the transmission before these reflections enter the measurement region. In practice, a stationary interference pattern is reached after several internal reflections and can be well-measured. A more precise result is obtained by taking the average of $T\left(t\right) $ for a short period of time in the final stationary region.

Note that although our initial wave packet is a good approximation of a plane wave, it also contains different wave vectors. This effect is mainly visible at the front and back of the wave packet and their contribution to the stationary interference pattern can be neglected. In the simulations for \emph{n-p-n} junctions the wave packet has a typical width of 50 wavelengths.

Until now, we have discussed the case when the potential is the same in the left and right measurement regions. When this is not the case, the incoming and transmitted waves have different group velocities along the $x$-direction. Therefore, one needs to correct Eq.~(\ref{eq:defTrans}) for this difference:
\begin{equation}
  T\left( t\right) = \frac{v_{g,out}}{v_{g,in}} \frac{A_{out}\left( t\right) }{A_{in}} = \frac{\cos\theta}{\cos\phi} \frac{A_{out}\left( t\right) }{A_{in}}, \label{eq:defTrans2}
\end{equation}
where the last equality is only valid when we are in the Dirac regime, and $\phi$ and $\theta$ are the angles that the incoming and outgoing waves make with the $x$-axis, i.e. $\cos\phi=k_{x,in}/|\mathbf{k}_{in}|$. A more rigorous version of this argument can be obtained by calculating the conserved current for the Dirac Hamiltonian~(\ref{eq:Dirac}), $j_x = \Psi^\dagger \sigma_x \Psi$, for the incoming and outgoing waves.\cite{Tudorovskiy13}  Although this method is suitable when we are inside the Dirac regime, it is not at all trivial to devise a similar method outside of this regime.

\subsubsection{n-p and n-n' junctions} \label{sec:method_nn}
\begin{figure}[tbp]
\begin{center}
\includegraphics[width=1.0\columnwidth]{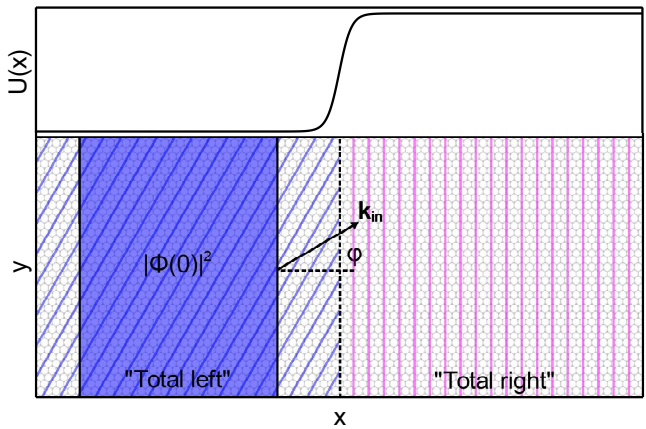}
\end{center}
\caption{Simulation setup for \emph{n-p} and \emph{n-n'} junctions. The junction is located at the center. The areas for which the density is calculated cover the whole lattice. After the whole wave packet is either reflected or transmitted at the junction, the reflection and transmission are obtained. To prevent interference at the borders, a spacing between the initial wave packet and the left edge is introduced.}
\label{fig:simulation_setup_nn}
\end{figure}

To determine the reflection and transmission for \emph{n-p} and \emph{n-n'} junctions we use an adjusted simulation setup, shown in Fig. \ref{fig:simulation_setup_nn}. In this setup, the sample is divided into two parts by the center of the potential ($x=0$). We call the sum of the wave density in the left region ($x<0$) ``total left'' and the sum of the wave density in the right region ($x>0$) ``total right''.

When the entire wave packet has interacted with the barrier, i.e. has been partially reflected and partially transmitted, we determine the reflection and transmission by reading out the total densities in the left and right region, respectively. One should note that to prevent interference from the reflections at the left and right boundaries of the sample, a spacing between the initial wave packet and the left border is necessary.

With this method, the problem with different group velocities for the incoming and reflected waves is circumvented and the transmission and reflection can be determined independently of the potential on the right side of the junction. The accuracy of the method depends on the size of the wave packet, since additional wave vectors are introduced due to the finite size. Naturally, their influence can be reduced by increasing the length of the initial wave packet. 
Note that this method is not able to deal with internal reflections and therefore cannot be used for \emph{n-p-n} junctions. On the other hand, the absence of internal reflections in \emph{n-n'} junctions enables us to use smaller samples. In the simulations of \emph{n-n'} junctions, the wave packet has a typical width of five wavelengths.

\section{Klein Tunneling} \label{sec:KT}

In general, \emph{n-p-n} and \emph{n-n'} junctions are quasi one-dimensional structures. In this paper, we will model them by a potential that only depends on the $x$-coordinate, $U=U(x)$. Because of this, the transversal wave vector $k_y$ is conserved.

\subsection{\emph{n-p-n} junction}

For a sharp rectangular \emph{n-p-n} junction, the jump in the electrostatic potential at the interface is given by a step function%
\begin{equation}
U(x)=\left\{
\begin{array}{rl}
U_0, & \quad 0\leq x\leq d \\
0, & \quad \text{otherwise}%
\end{array}%
\right. , \label{eq:pot_npn_rectangular}
\end{equation}%
where $d$ is the width of the potential barrier and $U_0$ the height of the barrier.
Within the Dirac approximation~(\ref{eq:Dirac}), the transmission for an electron with kinetic energy $E<U_0$ can be analytically calculated as
$T=1-|r|^2$, where\cite{katsnelson-nature-2006}
\begin{equation}
|r| = \frac{2 \sin(q_x d) (\sin\phi+\sin\theta)}{|e^{-i q_x d}\cos(\phi+\theta)+e^{i q_x d}\cos(\phi-\theta)+2 i \sin(q_x d)|} .
\label{eq:transmission_npn}
\end{equation}
In this expression, $\varphi$ is the incidence angle, $q_{x}=\sqrt{\left( E-U_0\right)^{2}/\hbar ^{2}v_{F}^{2}-k_{y}^{2}}$ is the $x$-component of the wave vector of the transmitted wave, and $\theta$ is the angle of the transmitted wave, defined by $E \sin\varphi=|E-U_0|\sin\theta$. The above equation shows that at normal incidence, i.e. $\varphi =0$, the reflection coefficient $r$ is zero,
the so-called Klein tunneling. Another feature of Eq.~(\ref{eq:transmission_npn}) is that there is total transmission whenever $q_x d$ is a multiple of $\pi$. The angles at which this occurs are called \emph{magic angles}.\cite{katsnelson-nature-2006}

\begin{figure}[tbp]
\centering
\mbox{
\includegraphics[width=1.0\columnwidth]{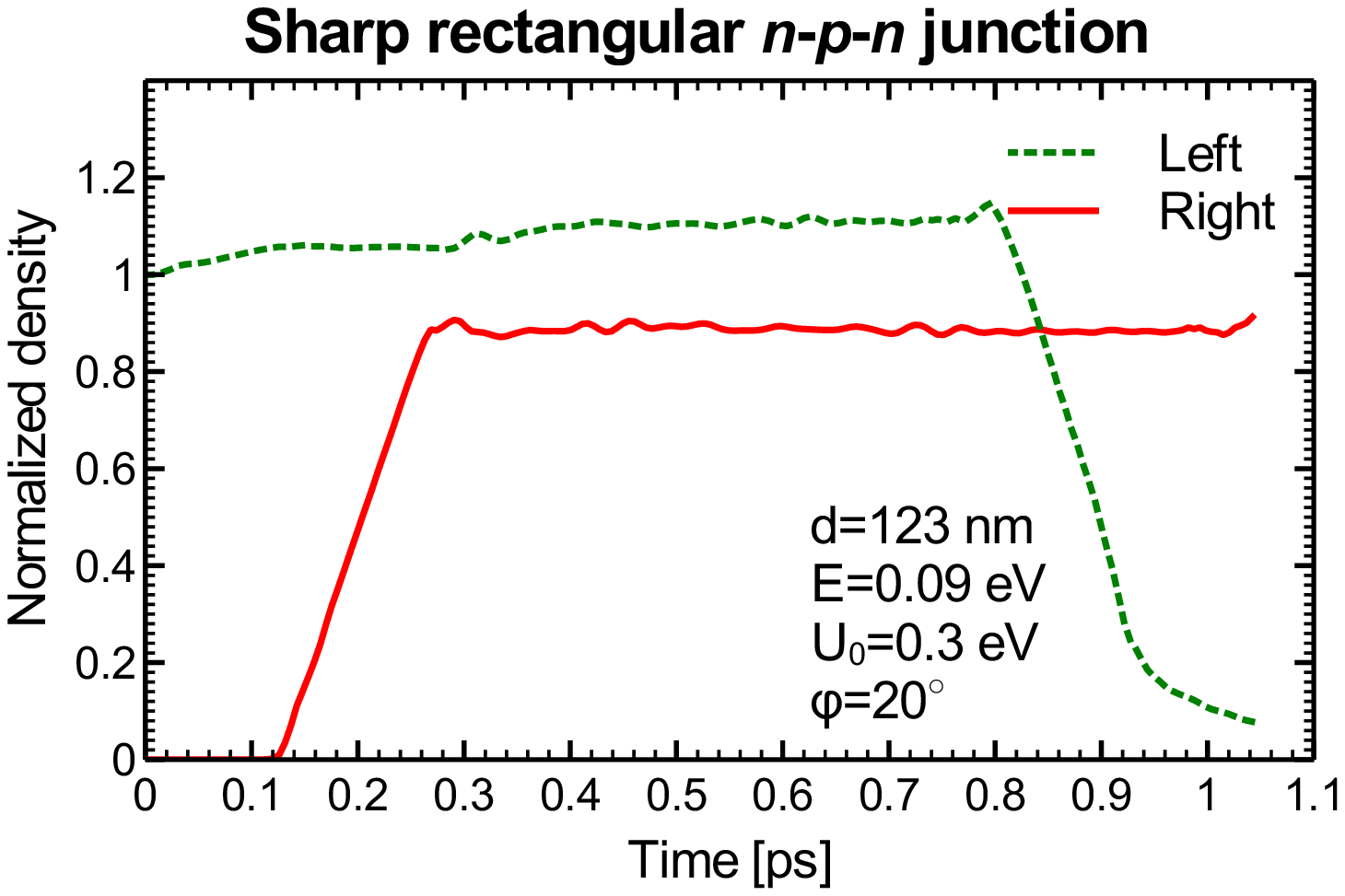}
}
\mbox{
\includegraphics[width=1.0\columnwidth]{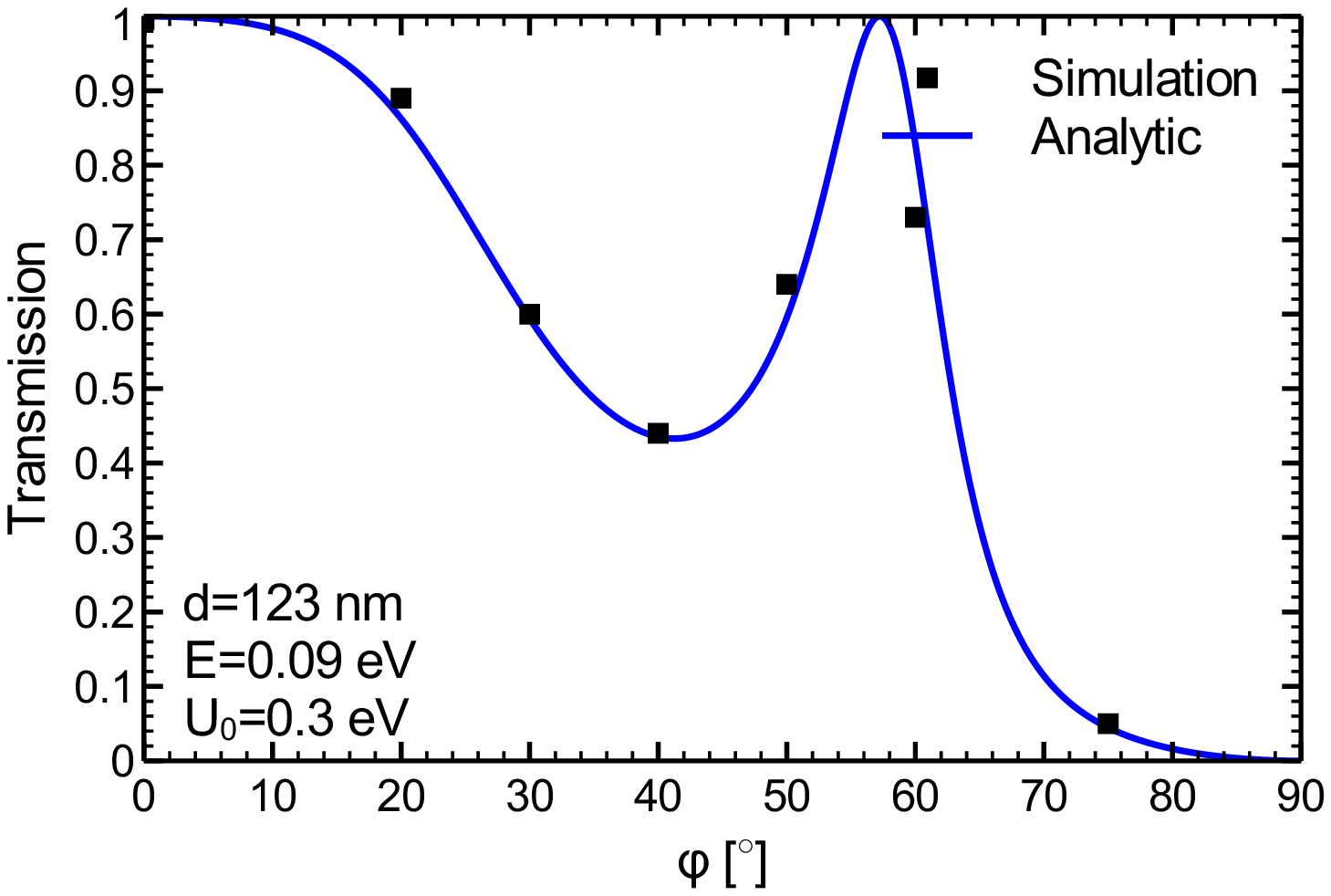}
}
\caption{Transmission for a sharp rectangular \emph{n-p-n} junction with $U_0=0.3$~eV, $E=0.09$~eV and $d=123$~nm. (Top) Normalized densities in the ``left'' (green dashed line) and ``right'' (solid red line) measurement regions (see Fig.~\ref{fig:simulation_setup_npn}) as a function of time, from which the transmission for the incidence angle $\varphi=20^\circ$ is extracted.
(Bottom) Transmission as a function of incidence angle $\varphi$. The numerical results agree very well with the analytic solution~(\ref{eq:transmission_npn}).}
\label{fig:rect_npn}
\end{figure}

In Fig.~\ref{fig:rect_npn} (top), we show the result of a simulation for a sharp rectangular \emph{n-p-n} junction. The transmission as a function of time is extracted using the method of Section~\ref{sec:method_npn}. When the wave packet enters the measurement region, the density increases approximately linearly, and after that it rapidly converges. In Fig.~\ref{fig:rect_npn} (bottom), the transmission through the junction is plotted as a function of incidence angle. We see that there is good agreement between the results of the numerical simulation and the analytical result~(\ref{eq:transmission_npn}).

For a more realistic model of an \emph{n-p-n} junction, one can consider a smooth potential, such as
\begin{equation}
U(x)=\tfrac{U_0}{2}\left[ \mathrm{tanh}\left( \tfrac{10x}{\ell_{1}}%
-5\right) -\mathrm{tanh}\left( \tfrac{10(x-\ell_{1}-\ell_{2})}{\ell_{3}}-5\right) %
\right] , \label{eq:pot_npn_smooth}
\end{equation}%
where $U_0$ is the maximum of the potential, $\ell_{2}$ is the length
of the barrier plateau and $\ell_{1}$ and $\ell_{3}$ are the typical distances of the potential increase and decrease, respectively.

\begin{figure}[tbp]
\begin{center}
\mbox{
\includegraphics[width=1.0\columnwidth]{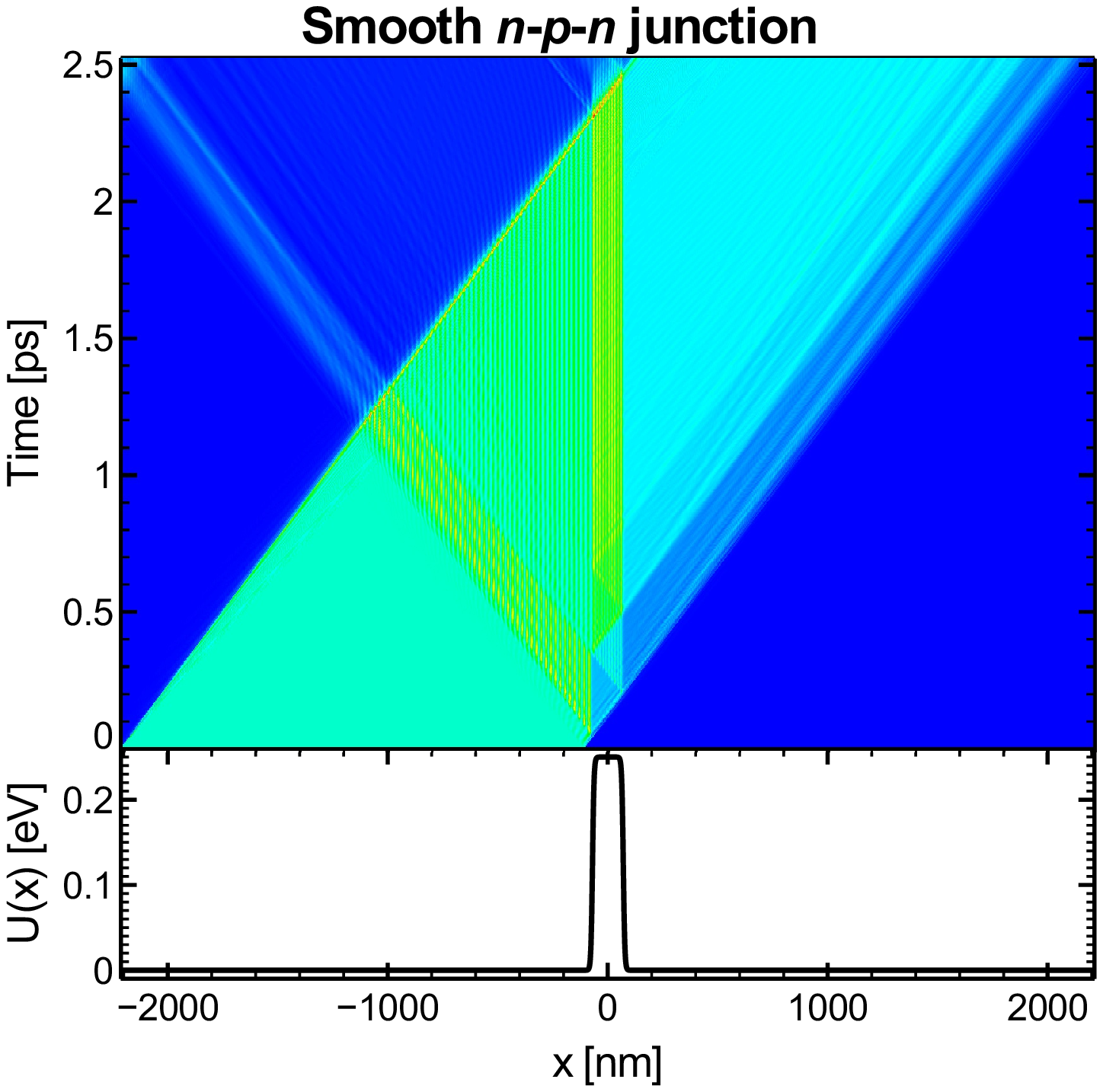}
}
\mbox{
\includegraphics[width=1.0\columnwidth]{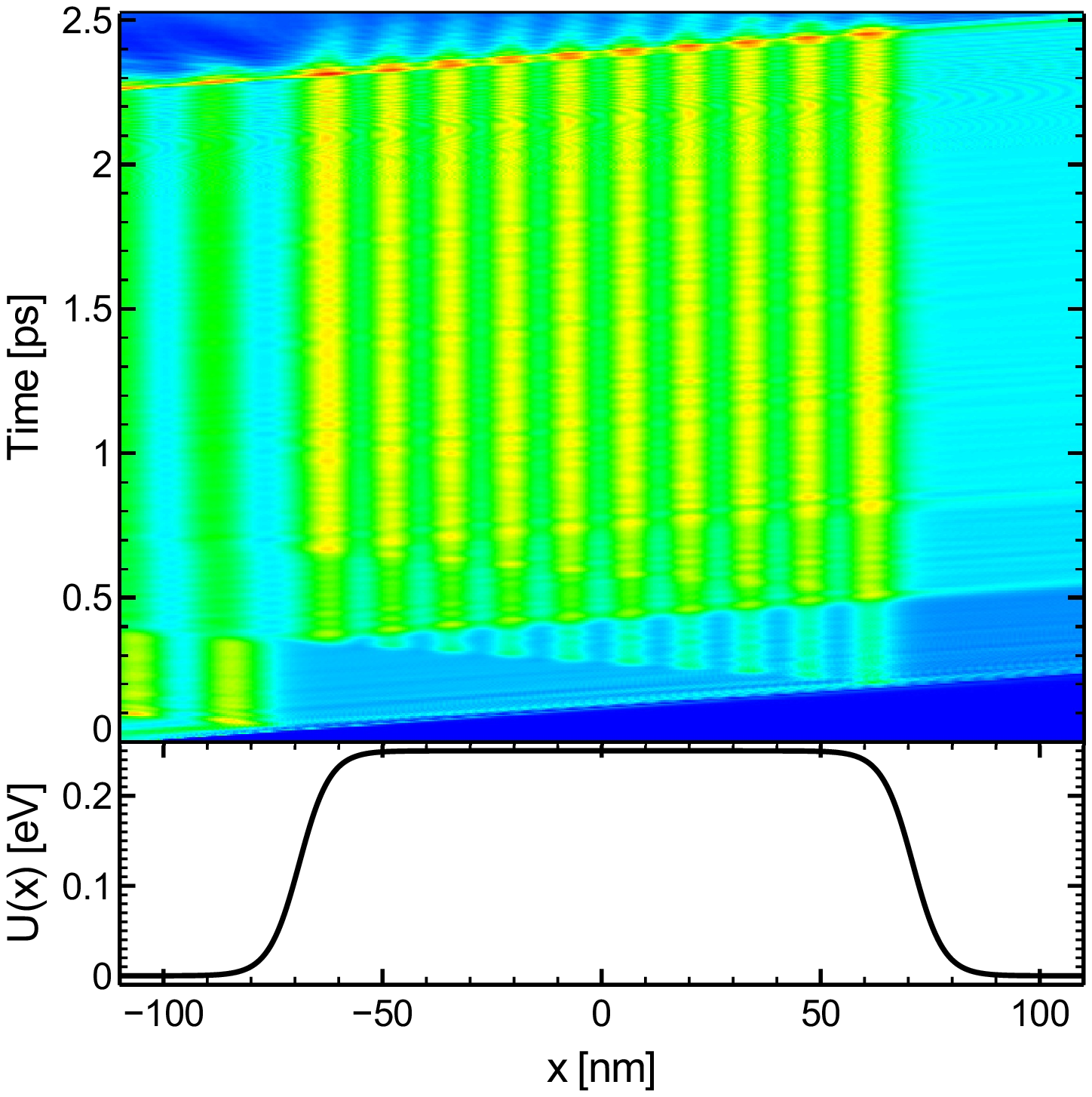}
}
\end{center}
\caption{Top: The evolution of the density, integrated along the $y$-direction, for every $x$ as a function of time for the smooth \emph{n-p-n} junction at $\protect\varphi$=20$^{\circ}$. Blue and red indicate low and high densities, respectively. Bottom: Zoom of the junction area. Note the internal reflections within the barrier, which are converged after three full reflections.}
\label{fig:movie_npn_smooth}
\end{figure}

We can compare the results of our numerical simulations with analytical results that where obtained using the semiclassical approximation.\cite{Shytov08,Tudorovskiy13,Reijnders14} The accuracy of this approximation is controlled by the (dimensionless) semiclassical parameter $h$, defined by $h=\hbar/p_0 l$, where $l$ is the intrinsic length scale of the problem, i.e. the typical scale of a change in the potential, and $v_F p_0$ is the characteristic value of $|U(x)-E|$. Put differently, $h$ is simply the ratio of the typical de Broglie wavelength $\hbar/p_0$ and the typical length scale $l$. The accuracy of the approximation increases when $h$ decreases.

Within the semiclassical approximation, the transmission through an \emph{n-p-n} junction can be calculated as an infinite sum over internal reflections, \cite{Shytov08,Tudorovskiy13,Reijnders14}
\begin{align}
T_{tot} &= \left| \frac{t_{np\rightarrow }t_{pn\rightarrow }e^{\tfrac{-i S}{h}}}{%
1-r_{np\leftarrow }r_{pn\rightarrow }e^{\tfrac{-2 i S}{h}}} \right|^2  \nonumber
\\
&= \left| t_{np\rightarrow }t_{pn\rightarrow }e^{\tfrac{-i S}{h}}\sum_{n=0}^{\infty
}\left( r_{np\leftarrow }r_{pn\rightarrow }e^{\tfrac{-2 i S}{h}}\right) ^{n}\right|^2. \label{eq:seminpn}
\end{align}%
In this expression, $t_{np\rightarrow }$ and $r_{np\rightarrow }$ are the transmission and reflection coefficients for an \emph{n-p} junction with an incident wave from the left, and the other quantities are named in a similar fashion.
Furthermore, $S$ is the semiclassical action inside the barrier,
\begin{equation}
  S = \frac{1}{p_0 l} \int_{x_-}^{x_+} \sqrt{(U(x)-E)^2/v_F^2-p_y^2} \,\text{d}x ,
\end{equation}
where $x_\pm$ are the classical turning points, i.e. the roots of $(U(x)-E)^2/v_F^2-p_y^2$. The transmission and reflection coefficients in Eq.~(\ref{eq:seminpn}) are expressed in terms of the action $K$ in the classically forbidden region between the electron and hole regions, and both $K$ and $S$ can be calculated semi analytically; see Ref.~\onlinecite{Reijnders14}.

\begin{figure}[tbp]
\centering
\mbox{
\includegraphics[width=1.0\columnwidth]{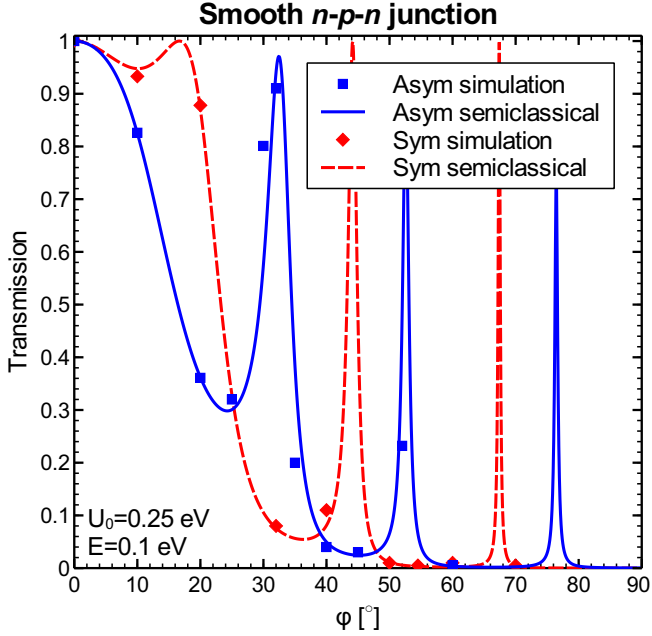}
}
\caption{Transmission of a wave packet with energy $E=0.1$~eV through a symmetric~(Sym, solid blue line) and an asymmetric~(Asym, dashed red line) smooth \emph{n-p-n} junction as a function of incidence angle $\protect\varphi$. For the symmetric potential $\ell_{1}=\ell_{2}=\ell_{3}=70$~nm, whereas for the asymmetric potential $\ell_{1}=50$~nm, $\ell_{2}=100$~nm and $\ell_{3}=70$~nm. Both potentials have the same height $U_0=0.25$~eV, and the semiclassical parameter $h=0.09$.
The agreement between the numerical results and the semiclassical solution~(\ref{eq:seminpn}) is very good.
}
\label{fig:smooth_npn_trans}
\end{figure}

In Fig.~\ref{fig:movie_npn_smooth}, we show the time evolution of the wave packet in our numerical simulations, for a typical smooth \emph{n-p-n} junction. As before, we have plotted the density, integrated along the $y$-direction, for every point $x$ at each time $t$. One can clearly see that the density inside the barrier increases in time, and that for this angle the stationary pattern is reached after three full internal reflections. 

For the smooth \emph{n-p-n} junction~(\ref{eq:pot_npn_smooth}), we consider two different types of potential profiles: a symmetric junction with $\ell_{1}=\ell_{3}$ and an asymmetric junction with $\ell_{1}\neq \ell_{3}$. Figure~\ref{fig:smooth_npn_trans} shows a comparison of our numerical simulation and the semiclassical result~(\ref{eq:seminpn}), for both a symmetric ($\ell_{1}=\ell_{2}=\ell_{3}=70$~nm) and an asymmetric ($\ell_{1}=50$~nm, $\ell_{2}=100$~nm and $\ell_{3}=70$~nm) junction. The height of the potential barrier is fixed at $U_0=0.25$~eV and the energy of the incident electron is $E=0.1$~eV. We once again see very good agreement between the simulations and theoretical predictions. 
Note that for large incidence angles, we have no simulation results at the transmission peaks, seen in the semiclassical prediction. The first reason for this is that the peaks are very narrow and since the semiclassical result is an approximation, they can easily be missed. Second, an analysis of the semiclassical transmission~(\ref{eq:seminpn}) shows that for larger incidence angles more internal reflections are needed to reach numerical convergence, especially at the transmission maxima. This requires considerably longer wavepackets and hence much larger samples. Outside the transmission maxima this is not the case and the agreement is still very good.

\subsection{\emph{n-n'} junction}

A sharp \emph{n-n'} junction can be described by the step potential
\begin{equation}
U(x)=\left\{
\begin{array}{rl}
0, & \quad x\leq 0 \\
U_0, & \quad x>0
\end{array}
\right. ,  \label{eq:pot_nn_rectangular}
\end{equation}
with $E>U_0$. As we did for an \emph{n-p-n} junction, we can introduce a smooth potential to get a more realistic model:
\begin{equation}
U(x)=\tfrac{U_0}{2}\left[ \mathrm{tanh}\left( \tfrac{10x}{\ell}-5\right) \right] ,  \label{eq:pot_pn_smooth}
\end{equation}
where $\ell$ is the typical distance of the potential increase.

\begin{figure}[tbp]
\centering
\mbox{
\includegraphics[width=1.0\columnwidth]{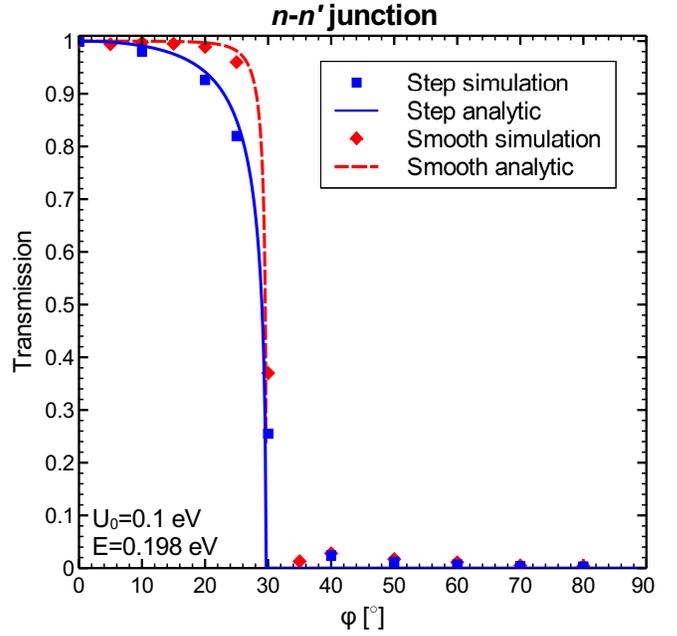}
}
\caption{Transmission of a wave packet with energy $E=0.198$~eV as a function of incidence angle $\protect\varphi$, for both a sharp (solid blue line) and smooth (red dashed line) \emph{n-n'} junction, which both have a height of $U_0=0.1$~eV. For the smooth junction $\ell=70$~nm, which corresponds to $h=0.09$. There is good agreement between the numerical and analytical results.}
\label{fig:nn_low}
\end{figure}

For electrons in graphene, the classical momentum $p_x(x)$ is given by
\begin{equation}
  p_x(x) = \sqrt{(U(x)-E)^2/v_F^2-p_y^2},
\end{equation}
where $v_F p_y=E \sin\varphi $, with $\varphi$ the angle of incidence. For an \emph{n-n'} junction, this means that the momentum at the right-hand side of the junction is imaginary whenever
\begin{equation}
  \sin \varphi > (U_0-E)/E, \label{eq:nnpr_forbidden}
\end{equation}
giving rise to a classically forbidden region. Therefore, we expect an exponentially decaying wave function in this region, instead of a plane wave. For angles $\phi$ that do not satisfy Eq.~(\ref{eq:nnpr_forbidden}), one can obtain an analytic solution for the reflection and transmission by matching waves at the barrier interface. Using more elaborate methods, one can also obtain a semiclassical result for the transmission for the smooth \emph{n-n'} junction~(\ref{eq:pot_pn_smooth}), see Ref.~\onlinecite{Reijnders14}.

In Fig.~\ref{fig:nn_low}, we show the simulated angle dependence of the transmission, for both a sharp and a smooth \emph{n-n'} junction, and compare it to the analytical results mentioned before. For both junctions the potential height is fixed at $U_0=0.1$~eV and the energy of the incident wave is fixed at $E=0.198$~eV. One sees that the agreement between numerical and analytical results is very good, and that the smooth \emph{n-n'} junction generally has a higher transmission than the sharp \emph{n-n'} junction with the same potential height.
The transmission in Fig.~\ref{fig:nn_low} has been extracted using the method outlined in Section~\ref{sec:method_nn}. For the sharp potential step, we have checked that, for incidence angles $\varphi$ that do not satisfy Eq.~(\ref{eq:nnpr_forbidden}), the same results can be obtained by using the method from Section~\ref{sec:method_npn} when we use Eq.~(\ref{eq:defTrans2}) to extract the transmission. However, the method from Section~\ref{sec:method_nn} allows us to use smaller samples.
Note that the transmission for angles that satisfy Eq.~(\ref{eq:nnpr_forbidden}) is not equal to zero. 
We attribute this to other $k$-vectors that are present in the wave packet. 
For wave vectors with larger $k$, the area to the right of the barrier is not forbidden, whereby they give rise to propagating waves and hence to nonzero transmission.

\subsection{Beyond the Dirac regime}

Since we use the tight-binding model in our numerical simulations, we can also study electron wave propagation beyond the Dirac cone approximation. To see whether Klein tunneling persists beyond the Dirac regime, one can consider the case of a weak potential and consider the transition matrix element in the first order Born approximation,
\begin{align}
  T^{(1)}(\mathbf{k}',\mathbf{k}) &= \langle \mathbf{k}' | U(\mathbf{x}) | \mathbf{k} \rangle \nonumber \\
  &= \frac{U_{\mathbf{k'}-\mathbf{k}}}{2} \left( 1 + \frac{\lambda_1 f^*(\mathbf{k'})}{|f(\mathbf{k'})|} \frac{\lambda_2 f(\mathbf{k})}{f(\mathbf{k})|} \right), 
  \label{eq:TMatrixExp}
\end{align}
where $U_{\mathbf{k'}-\mathbf{k}}$ represents a Fourier component of the potential $U(\mathbf{x})$, and we have used Eq.~(\ref{Eq:WaveK}). The constant $\lambda_1$ ($\lambda_2$) equals $\pm1$, depending on whether the state $|\mathbf{k'}\rangle$ ($|\mathbf{k}\rangle$) is an electron or a hole state. In the first order Born approximation, the probability of backscattering from an inital state $|\mathbf{k}_\text{in}\rangle$ to a final state $|\mathbf{k}_\text{back}\rangle$ is proportional to $|T^{(1)}(\mathbf{k}_{\text{back}},\mathbf{k}_{\text{in}})|^2$. So if the matrix element~$T^{(1)}(\mathbf{k}_{\text{back}},\mathbf{k}_{\text{in}})$ is nonzero, then backscattering is allowed and there is no Klein tunneling. Note that since the potential $U$ is scalar, that is, just proportional to the unit matrix in pseudospin space, this only happens whenever the wave functions $|\mathbf{k}_\text{in}\rangle$ and $|\mathbf{k}_\text{back}\rangle$ are orthogonal in pseudospin space. In the Dirac regime this is indeed the case, since $\mathbf{k}_\text{back} = -\mathbf{k}_\text{in}$, and the term $\frac{f(\mathbf{k})}{f(\mathbf{k})|}$ equals minus one for the incoming state and plus one for the scattered state; see, e.g., Ref.~\onlinecite{Allain11}. However, it is important to understand that the vanishing of the matrix element~$T^{(1)}(\mathbf{k}_{\text{back}},\mathbf{k}_{\text{in}})$ does not guarantee Klein tunneling, since higher order terms in the Born series may not vanish and hence allow backscattering. Therefore, additional considerations are required in this case, such as a more detailed analysis that includes higher order terms in perturbation theory,~\cite{Ando98,KatsnelsonBook} or arguments based on pseudospin conservation.~\cite{katsnelson-nature-2006,Allain11,Tudorovskiy13}

Let us now investigate backscattering beyond the Dirac regime. Since in this regime the energy $E(\mathbf{k})$ is no longer invariant under arbitrary rotations in momentum space, we consider a one-dimensional potential barrier that is directed under an angle $\alpha$ with the $Ox$ axis. This means that when we introduce a new coordinate system $(x',y')$  by rotating the original coordinate system $(x,y)$ by an angle $\alpha$, the potential $U(x')$ only depends on $x'$. One can then define ``normal incidence'' in two different ways. In the first definition, we demand that the transversal momentum $k_y'$ in the rotated coordinate system vanishes. In the second definition, we require the group velocity, $\mathbf{v}_g(\mathbf{k})=\partial E(\mathbf{k})/\partial \mathbf{k}$, with $E(\mathbf{k})$ given by Eq.~(\ref{Eq:EnergyK}), to be orthogonal to the barrier interface. For general angles $\alpha$, these two definitions do not give the same momenta. However, for $\alpha =n\pi/3$, where $n$ is an integer, the two definitions are equivalent.

Let us first consider the first defintion, i.e. we demand that the transversal momentum $k_y'$ vanishes. As a first approximation, we can include trigonal warping effects in the Hamiltonian, that is, we expand $f(\mathbf{k})$ from Eq.~(\ref{eq:energy_f}) to second order in $k_x$ and $k_y$ around the $\mathbf{K}$-point. This case was analyzed in detail in Ref.~\onlinecite{Ando98}. By solving for the momenta $k_{x,\text{in}}'$ of the incoming and $k_{x,\text{ref}}'$ of the reflected wave, the authors showed that for a generic angle $\alpha$ that is not a multiple of $\pi/3$ the matrix element $T^{(1)}(\mathbf{k}_{\text{back}},\mathbf{k}_{\text{in}})$, see Eq.~(\ref{eq:TMatrixExp}), does not vanish. Therefore, we conclude that the probability of backscattering is nonzero and that there is no Klein tunneling.

We have explored the second definition of normal incidence numerically, determining the $k$-vectors for which the group velocity is orthogonal to the barrier interface. Taking into account conservation of the transversal momentum $k_y'$ in the rotated coordinate system, we then obtained the wave vector of the reflected wave. Computing the associated wave functions, we find that they are not orthogonal in pseudospin space, and hence that the matrix element $T^{(1)}(\mathbf{k}_{\text{back}},\mathbf{k}_{\text{in}})$ does not vanish. Therefore, we conclude that there is no total transmission, just as in the first definition of normal incidence. However, note that for both definitions the overlap is fairly small. Therefore, the magnitude of the effect could be rather small, similar to the case where the next-nearest-neighbor hopping parameter $t'$ is included in the description.~\cite{Kretinin13}

\begin{figure}[tbp]
\centering
\includegraphics[width=1.0\columnwidth]{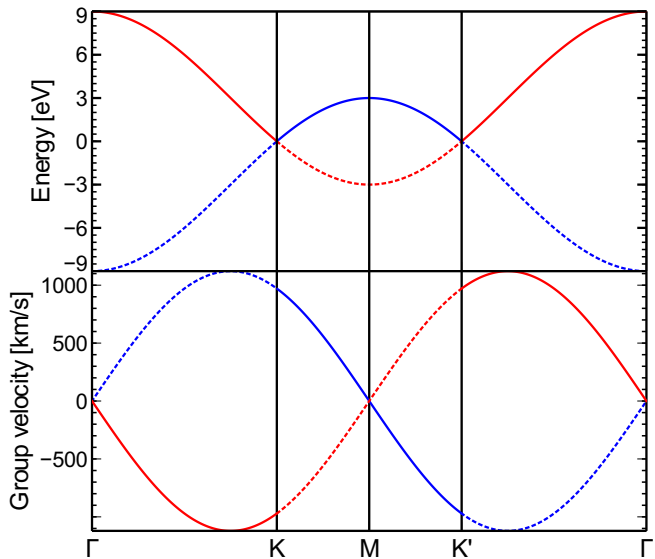}
\caption{The top graph shows the energy $E(k_x)$ for an electron with zero transversal momentum. The red line indicates that the spinor structure of the wave function is proportional to $(1,1)^T$ and the blue line that it is proportional to $({-1},1)^T$. Furthermore, a solid line indicates an electron, and a dashed line indicates a hole. The bottom graph shows the group velocity for the above particles, with the same color coding.}
\label{fig:energy_full}
\end{figure}

Because of the previous discussion, we will from now on consider the special case $\alpha =n\pi/3$, where Klein tunneling is not excluded by perturbative arguments. This corresponds to the samples we have studied numerically in the previous sections, that is, those with zigzag boundaries in the $x$-direction and armchair boundaries in the $y$-direction. Without loss of generality, let us consider $\alpha=0$, so that normal incidence corresponds to $k_y=0$. Then the function $f(\mathbf{k})$, given by Eq.~(\ref{eq:energy_f}), reduces to
\begin{equation}
  f(k_x) = 1 + 2 \cos(\sqrt{3}k_x a/2) .
  \label{eq:f_kx_normal}
\end{equation}
In the absence of a potential $U(x)$, the Hamiltonian~(\ref{Eq:SLG}) in momentum space then equals
\begin{equation}
  H(k_x)=t f(k_x) \sigma_x \label{eq:H_kx_normal} .
\end{equation}
Since the only Pauli matrix it contains is $\sigma_x$, this Hamiltonian can be exactly diagonalized and the wave functions are either proportional to $(1, 1)^T$, or to $({-1}, 1)^T$. This can also be seen from Eq.~(\ref{eq:coeffs_ai}), since one finds from Eq.~(\ref{eq:f_kx_normal}) that $f(k_x)$ is real. In Fig.~\ref{fig:energy_full}, we show the energy $E(k_x)$, given by Eq.~(\ref{Eq:EnergyK}), over the full Brillouin zone. The corresponding eigenvectors are indicated by using two colors, red for $(1, 1)^T$ and blue for $({-1}, 1)^T$. Since these wave functions are orthogonal in pseudospin space, the matrix element $T^{(1)}(k_{x}',k_{x})$ for scattering between these states vanishes. In appendix~\ref{app:Tmatrix}, we show that all higher order terms in perturbation theory also vanish. Therefore, scattering between a state that is proportional to $(1, 1)^T$ and a state that is proportional to $({-1}, 1)^T$ is forbidden. A weaker version of this statement was proven in Ref.~\onlinecite{Ando98}, where the authors only considered scattering within a single valley, although the generalization is straightforward. In the appendix, we do not follow their detailed considerations, but instead present a simplified version of the argument, similar to the discussion in Ref.~\onlinecite{KatsnelsonBook}, which is sufficient for one-dimensional scattering.

\begin{figure}[tbp]
\centering
\includegraphics[width=1.0\columnwidth]{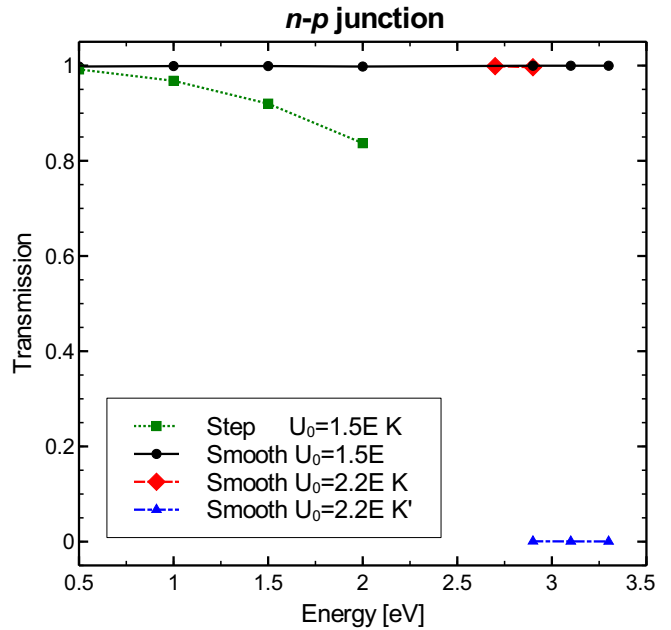}
\caption{Transmission through an $n$-$p$ junction for energies $E$ that are outside of the Dirac regime. The green squares show the result for a sharp potential step, whereas all the other points have been obtained with the smooth potential~(\ref{eq:pot_pn_smooth}), with $\ell=70$~nm. For the sharp potential one sees a decay in the transmission due to intervalley scattering. For the smooth potential there is (almost) total transmission when there are hole states with the same spinor structure. Otherwise there is no transmission, or total reflection.}
\label{fig:np_regimes_high}
\end{figure}

Let us now investigate which types of scattering processes there are outside of the Dirac regime. At the $M$-point there is a Van Hove singularity, where the energy is $\pm t$, (see Fig.~\ref{fig:energy_full}). Since both $E$ and $U_0-E$ can be smaller or larger than $t$, we identify four different scattering regimes. In Fig.~\ref{fig:np_regimes_high}, we show the simulation results for all these different regimes, where the transmission has been extracted using the method from Sec.~\ref{sec:method_nn}. Let us first concentrate on the first scattering regime, where both $E$ and $U_0-E$ are smaller than $t$. For a sharp potential barrier~(\ref{eq:pot_nn_rectangular}), we see that the transmission is no longer equal to one and that it decays as a function of the energy of the incoming electron. When we look at Fig.~\ref{fig:energy_full}, we see that the finite probability of backscattering is due to intervalley scattering: an incoming electron with a wave vector to the left of the $M$-point (it is closest to $\mathbf{K}$) is scattered to a reflected electron state with a wave vector to the right of the $M$-point. Such processes are allowed, since both states have the same structure in pseudospin space; they are proportional to $({-1}, 1)^T$.
Since the Fourier components $U_{\mathbf{k}-\mathbf{k}'}$ decay as a function of $|\mathbf{k}-\mathbf{k}'| \ell$, $\ell$ being the spatial scale of the potential, this intervalley scattering can be strongly suppressed by considering a smooth barrier~(\ref{eq:pot_pn_smooth}), with a sufficiently large value of $\ell$. In Fig.~\ref{fig:np_regimes_high}, we see that for $\ell=70$~nm, which means that $|\mathbf{k}-\mathbf{k}'| \ell$ is of the order of $10^2$, intervalley scattering is strongly suppressed, and we find that there is (almost) total transmission. We have also observed almost total tranmission for $\ell=10$~nm, which corresponds to the smaller value $|\mathbf{k}-\mathbf{k}'| \ell \sim 10^1$.  

When the energy $E$ becomes larger than $t$, the wave vector of the incoming electron is to the right of $\mathbf{K}'$. As long as $U_0-E<t$, one sees from Fig.~\ref{fig:energy_full} that the incoming electron can be scattered to a hole state with the same spinor structure as the incoming electron. Our numerical simulations for a smooth barrier show that there is (almost) total transmission in this case. This situation changes drastically when both $E>t$ and $U_0-E>t$, since the available electron and hole states now have a different spinor structure. Since our theoretical analysis showed that scattering between the two different spinor structures is impossible, we expect zero transmission in this case, which is confirmed by our numerical simulations. Although only the smooth barrier is shown in Fig.~\ref{fig:np_regimes_high}, we have checked that the same result holds for a sharp barrier. When $E<t$ and $U_0-E>t$, the transmission strongly depends on the wave vector of the incoming electron, as can be seen by comparing the red and blue lines in Fig.~\ref{fig:np_regimes_high} at $E=2.9$~eV. For an incoming electron with a wave vector that is closest to $\mathbf{K}'$, there are no hole states with the same spinor structure to which the electron can scatter, and our numerical simulations for a smooth barrier indeed show that there is zero transmission. For an electron that is closer to $\mathbf{K}$, such states are available, and our numerical simulations for a smooth barrier again show that there is (almost) unit transmission.

\section{Focusing by 2D potentials} \label{sec:caust}

The tight-binding propagation method is not limited to the study of one-dimensional potentials. In this section, we consider scattering by two-dimensional potentials with a maximum that is lower than the energy of the wave packet. Such potentials give rise to interference phenomena, and have the ability to focus the wave packet. This creates a possible way to control the propagation of electrons by introducing an effective \textit{optical} lens in graphene.

As an example of a potential that exhibits focusing, we consider a spherically symmetric Gaussian potential,
\begin{equation}
U(\mathbf{x})=\pm U_0 \,e^{-|\mathbf{x}-\mathbf{x_{0}}|^2/\ell^2}.  \label{eq:pot_gauss}
\end{equation}
where $\mathbf{x}_{0}$ is the center of the potential, and $\ell$ determines how fast it decays and thereby its width. Depending on its sign, this potential represents either a barrier ($+$) or a valley ($-$).

\begin{figure}[tbp]
\begin{center}
\includegraphics[width=1.0\columnwidth]{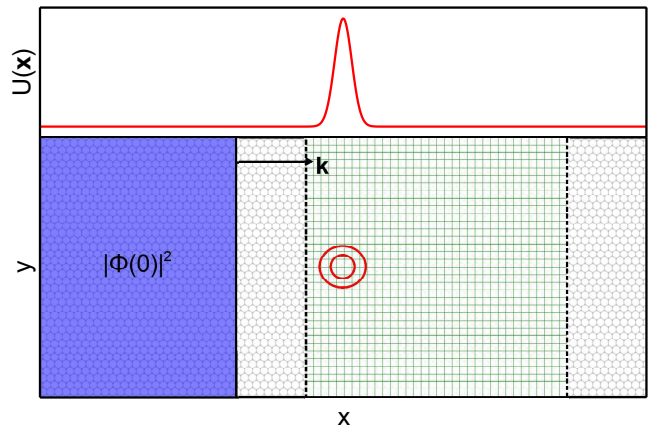}
\end{center}
\caption{Setup of the simulation. The Gaussian potential is located at the center of
the lattice and is indicated by two red circles. The initial wave packet is localized
on the left of the lattice and is indicated in blue. The wave propagation is
stored for the green (squared) area. }
\label{fig:simulation_setup_2d}
\end{figure}

The simulation setup is shown in Fig.~\ref{fig:simulation_setup_2d} and is quite similar to the one used for an \emph{n-p-n} junction. The Gaussian potential is located at the center of the sample, with the initial plane wave packet to its left. In the simulation, the plane wave packet propagates according to the TDSE and the wave density in the green area in Fig.~\ref{fig:simulation_setup_2d} is recorded. In order to reduce the required amount of storage, the wave density is averaged over blocks of $5\times 5$ atoms. The simulation is stopped when a stable interference pattern is reached.

\subsection{Classical electron trajectories}

We can compare the outcome of our simulations to the classical electron trajectories. These are similar to the rays in geometrical optics, and show where focusing takes place. Since this is a classical description, we expect to find good agreement only when the typical de Broglie wavelength of the electrons is much smaller than the typical length scale introduced by the potential. This means that the parameter $h$, introduced above Eq.~(\ref{eq:seminpn}), should be small.

To find the classical Hamiltonian for electrons, one should first introduce dimensionless parameters in Eq.~(\ref{eq:Dirac}), as was done in Ref.~\onlinecite{Tudorovskiy13}. One can then extract the classical Hamiltonians that are contained within the matrix Hamiltonian by replacing the operators $\hat{p}_x$ and $\hat{p}_y$ by the numbers $p_x$ and $p_y$ and computing the eigenvalues. This procedure gives two classical Hamiltonians, one for electrons and one for holes. For electrons, we find that
\begin{equation}
  H(\mathbf{p}, \mathbf{x})= v_F |\mathbf{p}| + U(\mathbf{x}) .
\end{equation}
In the problem under consideration, the potential $U(\mathbf{x})$ is given by Eq.~(\ref{eq:pot_gauss}). The trajectories $\mathbf{x}(t)$ can then be found from Hamilton's equations,
\begin{equation}
\dot{\mathbf{x}}=\frac{\partial H}{\partial \mathbf{p}} \quad \mathrm{and}%
\quad \dot{\mathbf{p}}=-\frac{\partial H}{\partial \mathbf{x}},
\end{equation}
which can be integrated numerically for any energy $E$.

\begin{figure}[tbp]
\begin{center}
\includegraphics[width=0.48\columnwidth]{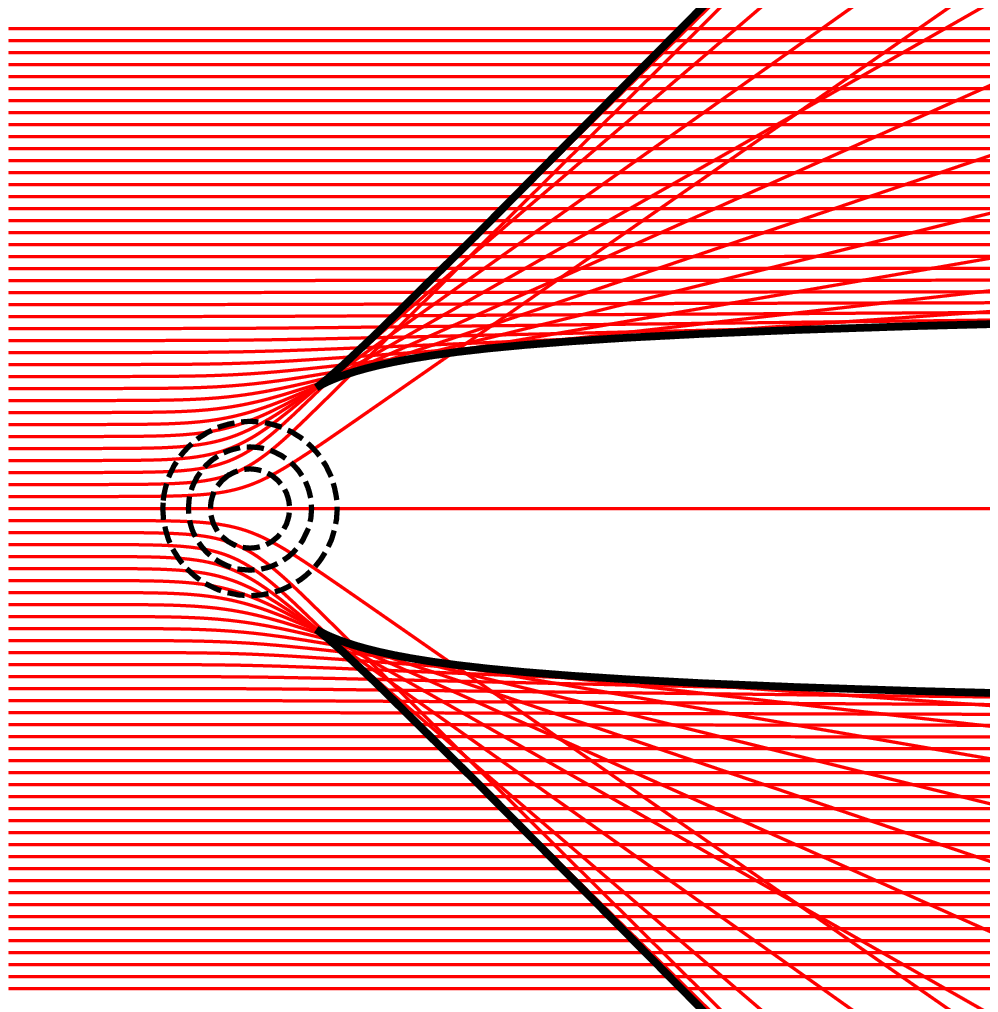}
\includegraphics[width=0.48\columnwidth]{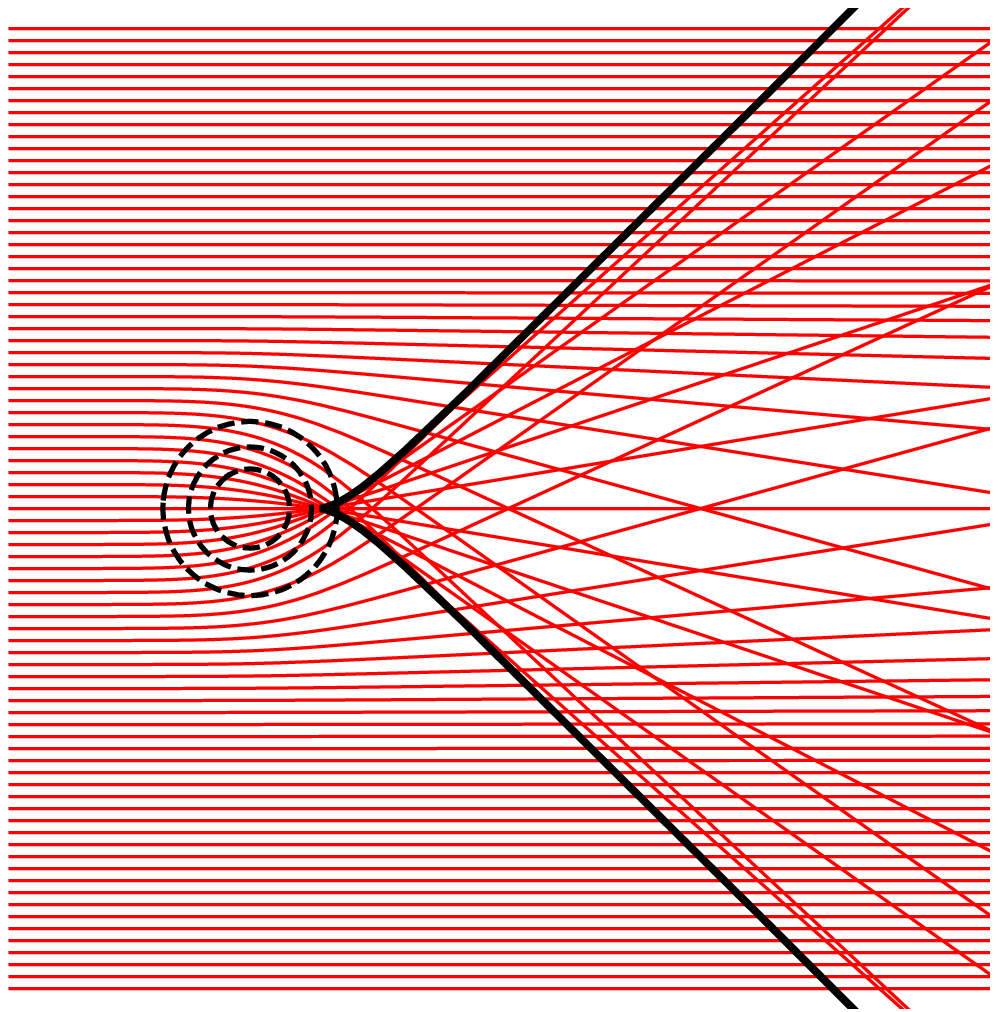}
\end{center}
\caption{The classical electron trajectories (solid red lines) for both a potential barrier (left) and a potential valley (right). The dashed lines indicate where the potential has decreased to 25, 50 and 75 percent of its maximum. A thick black line indicates a caustic, i.e. the envelope of the classical trajectories.}
\label{fig:trajectories}
\end{figure}

In Fig.~\ref{fig:trajectories}, we show the electron trajectories for both a potential barrier, for which the sign in Eq.~(\ref{eq:pot_gauss}) is positive, and a potential valley, for which the sign is negative. For both cases, the energy $E=0.198$~eV and the potential height $U_0=0.1$~eV. Note that when we introduce dimensionless variables, the new coordinates equal $\tilde{\mathbf{x}}=\mathbf{x}/w$. Hence, the electron trajectories for different widths of the potential can be obtained by scaling. For both the potential barrier and the valley, we see that the classical trajectories have an envelope, known as a \emph{caustic},\cite{Berry80,Poston78,Arnold90} and shown in black. Inside the envelope there is interference, because each point lies on three electron trajectories. Furthermore, we expect the intensity to be higher in regions where the density of trajectories is higher. Therefore, we expect the intensity to be low in the region behind the potential barrier.

\subsection{2D wave propagation}

\begin{figure*}[tbp]
\begin{center}
\mbox{
\includegraphics[width=\columnwidth]{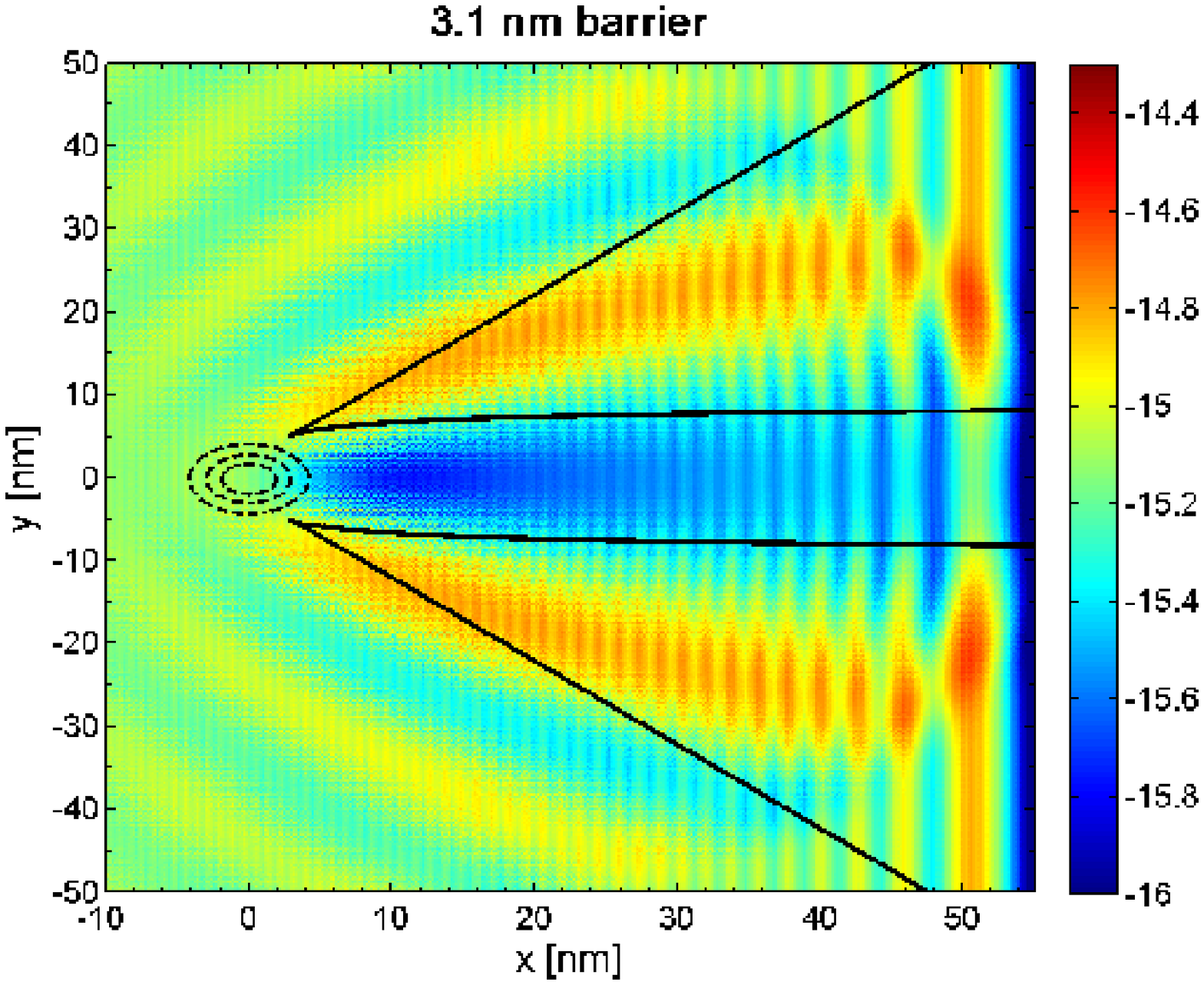}
\includegraphics[width=\columnwidth]{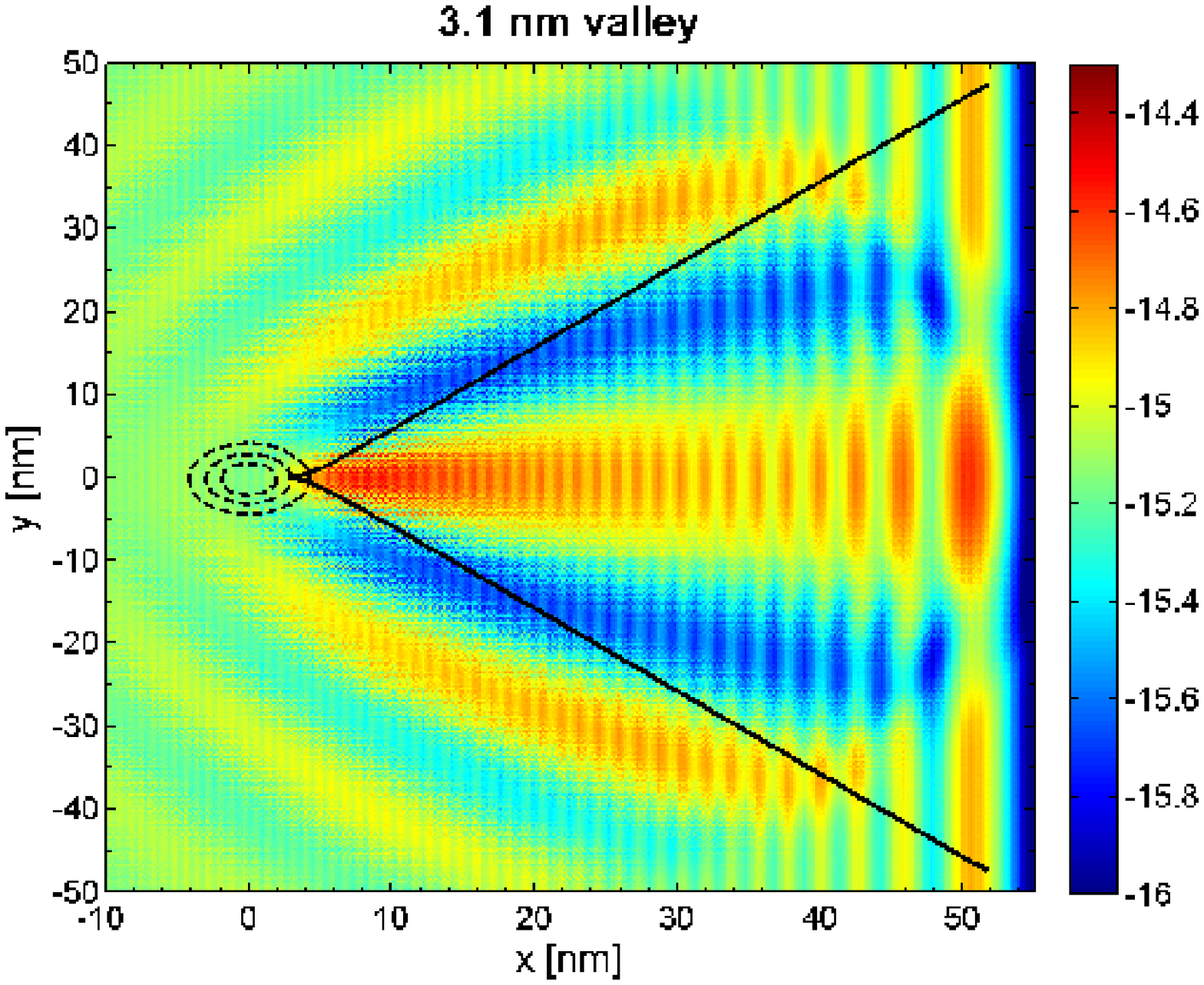}
}
\mbox{
\includegraphics[width=\columnwidth]{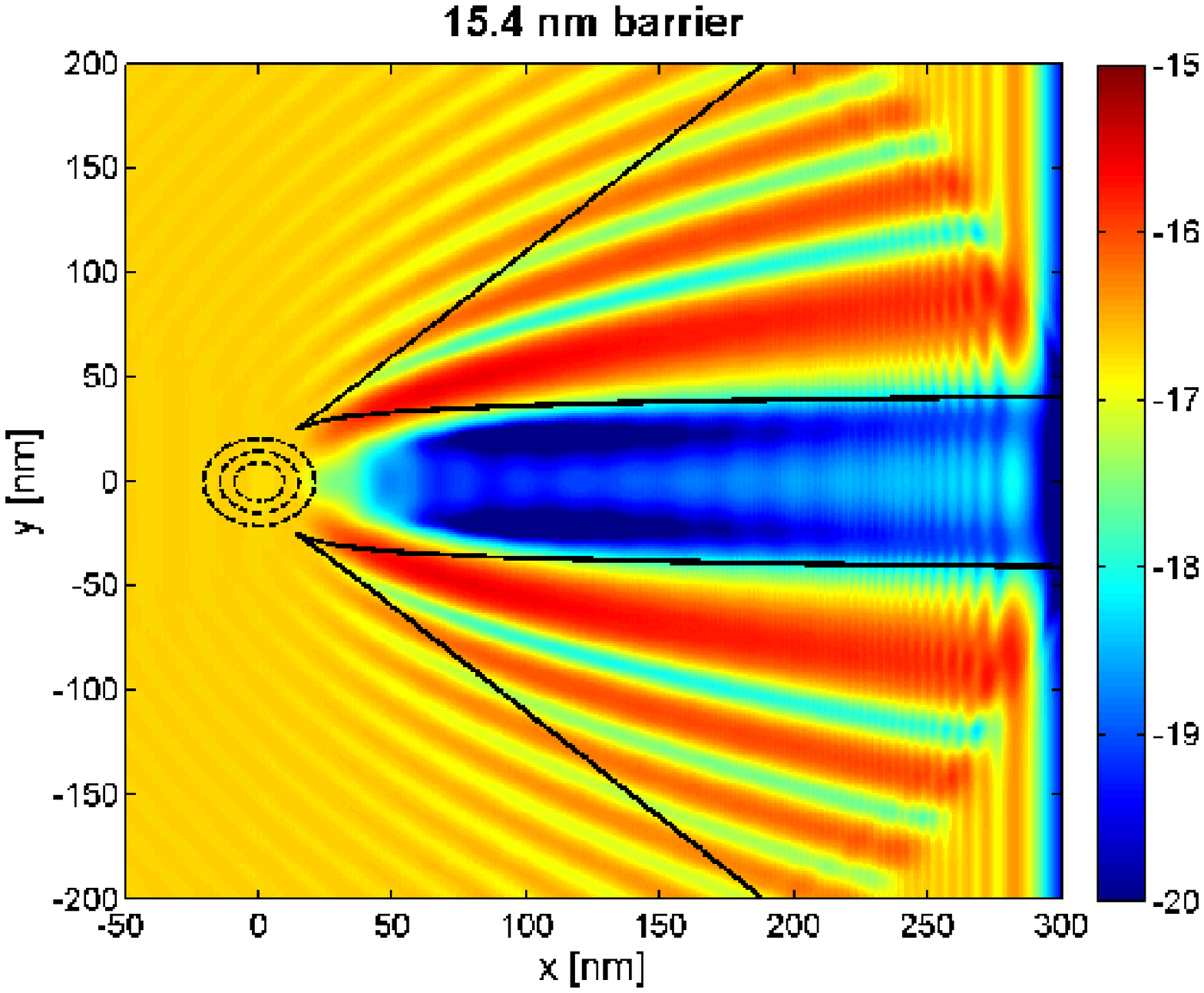}
\includegraphics[width=\columnwidth]{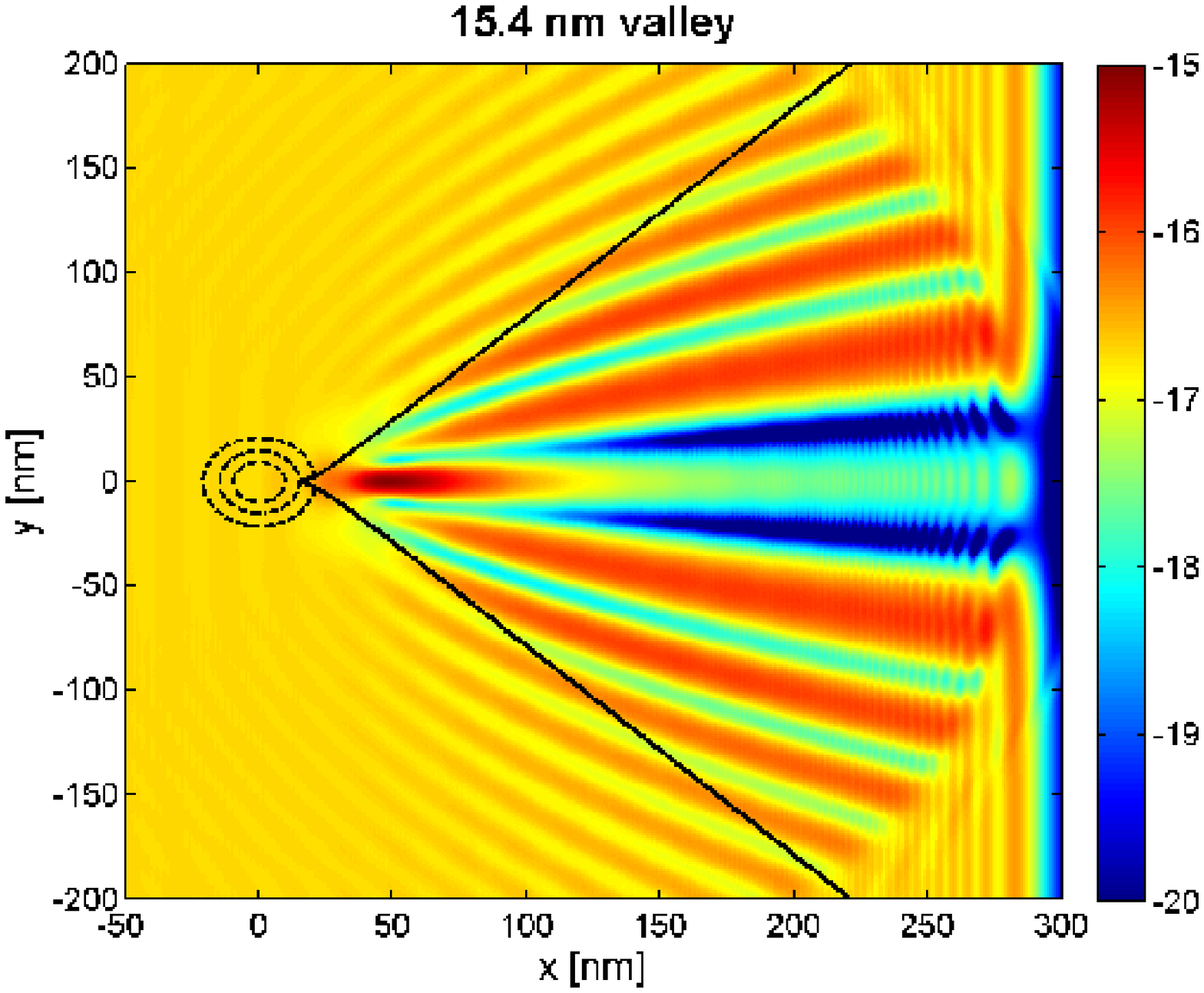}
}
\mbox{
\includegraphics[width=\columnwidth]{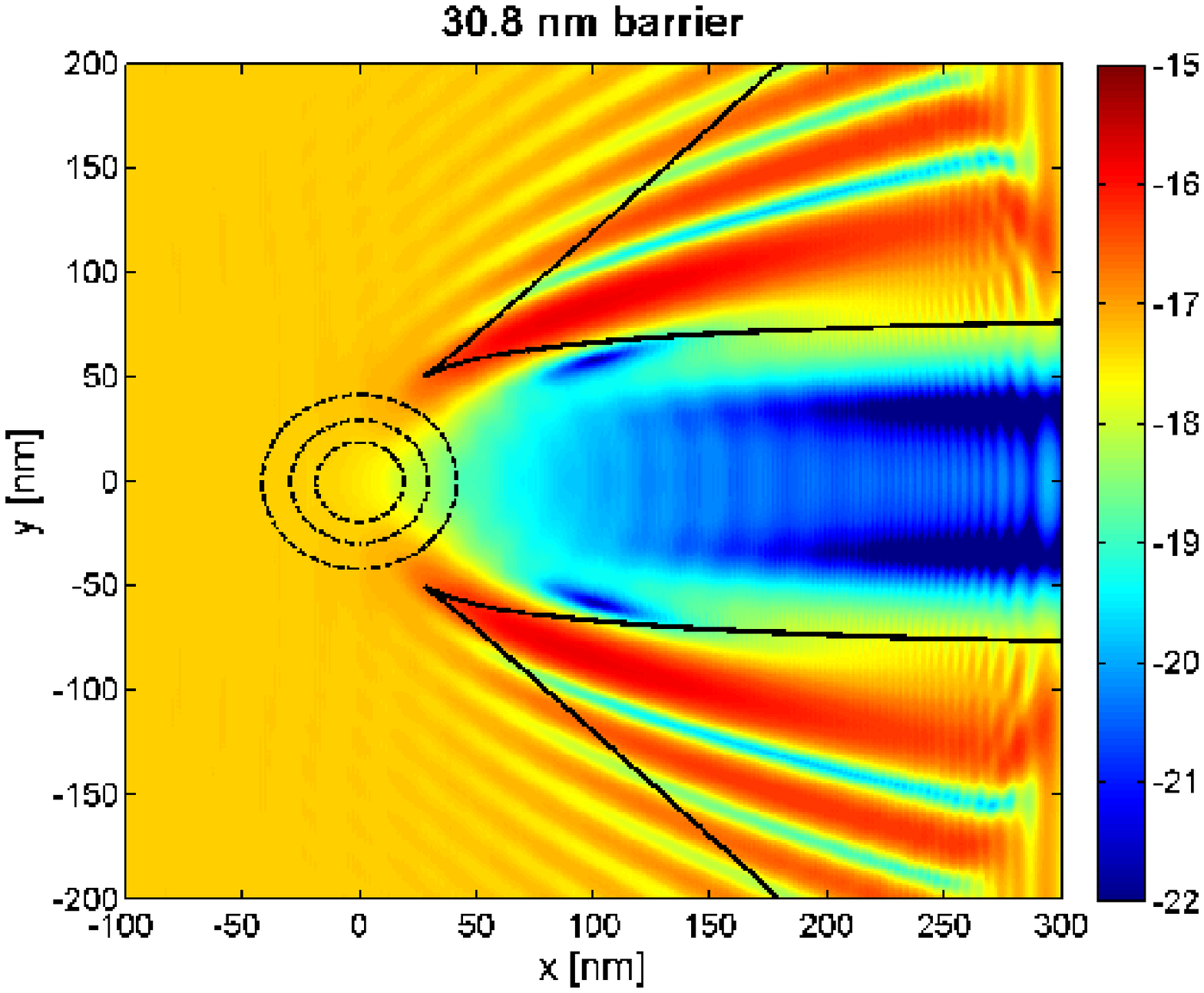}
\includegraphics[width=\columnwidth]{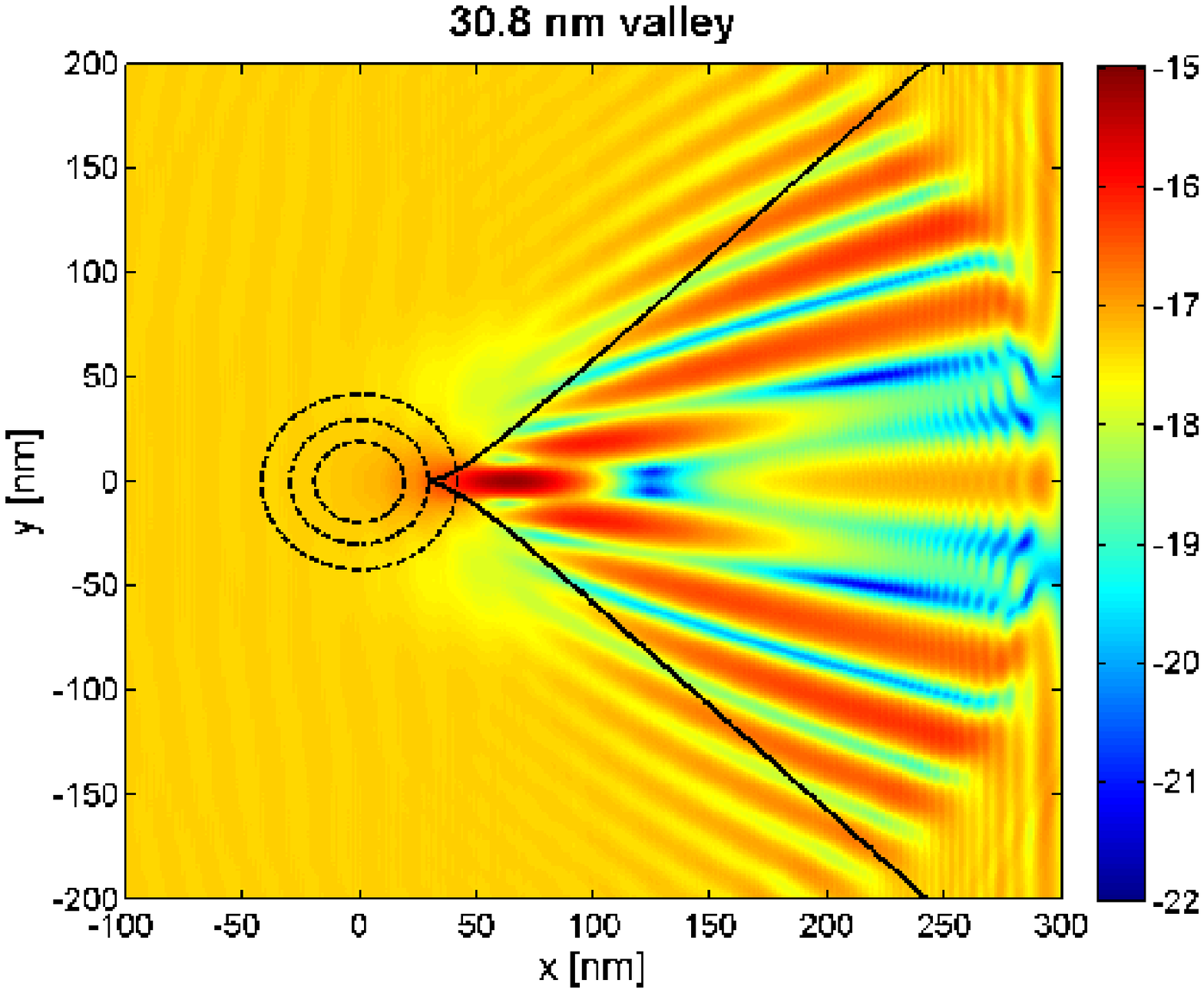}
}
\end{center}
\caption{Stable interference pattern for a wave packet with energy $E=0.198~eV$ incident on a Gaussian potential~(\ref{eq:pot_gauss}), with $U_0=0.1$~eV. Shown is the wave density in a logarithmic scale: $\log\left(|\Phi(\mathbf{x}) | \right)$. For the figures on the left, the sign of the potential is positive, corresponding to a barrier; on the right, the sign of the potential is negative, corresponding to a valley. Three different length scales are considered, $\ell=3.1$~nm, $\ell=15.4$~nm and $\ell=30.8$nm, corresponding to $h=2$, $h=0.4$ and $h=0.2$, respectively. As in Fig.~\ref{fig:trajectories}, the dashed lines indicate the contours of the potential, and a solid black line indicates a caustic. The agreement between the numerical simulation and the classical trajectories improves when the barrier becomes wider, i.e. when $\ell$ increases and $h$ decreases. 
}
\label{fig:2D_prop}
\end{figure*}

In Fig.~\ref{fig:2D_prop}, we show the stationary interference pattern for a wave packet with energy $E=0.198$~eV, incident on the potential~(\ref{eq:pot_gauss}), with $U_0=0.1$~eV. The figures on the left correspond to a potential barrier, and those on the right to a potential valley. The potential widths are determined by $\ell=3.1$~nm, $\ell=15.4$~nm and $\ell=30.8$~nm, corresponding to the semiclassical parameters $h=2$, $h=0.4$ and $h=0.2$, respectively.

These results can be compared with the classical electron trajectories (Fig.~\ref{fig:trajectories}) and the caustic, which is also shown in Fig.~\ref{fig:2D_prop}. For the smallest barrier, which is outside the semiclassical regime because of the large value of $h$, we see that the agreement is indeed very poor and that there is no real focus. When we increase the barrier width, we enter the semiclassical regime and the agreement indeed becomes much better. For both $\ell=15.4$~nm ($h=0.4$) and $\ell=30.8$~nm ($h=0.2$) we clearly see that the electrons are focused at the points predicted by the classical electron trajectories, with better agreement when $\ell=30.8$~nm. Furthermore, as predicted, we see a region of low intensity behind the potential barrier. For the potential valley with $\ell=30.8$~nm, we see the first interference maximum within the region bounded by the caustic.

\section{Conclusion} \label{sec:conclusion}

In this paper, we have studied Klein tunneling and quantum interference in graphene with the tight-binding propagation method. Using this numerical scheme, we have simulated the propagation of a plane wave packet according to the time-dependent Schr\"{o}dinger equation. Both sharp and smooth \emph{n-p-n} and \emph{n-n'} junctions have been considered, applying different methods to extract the transmission probability from the distribution of the wave function. In the case of an \emph{n-p-n} junction, quantum interference from multiple refections inside the barrier plays a crucial role. For an \emph{n-n'} junction, this problem does not exist, which allowed us to use smaller samples. Our results match very well to the analytical and semiclassical formulas applicable in the Dirac regime.

Since our numerical method is not restricted to this regime, we have also considered the transmission through an $n$-$p$ junction for energies outside the Dirac regime. We have found that when both $E<t$ and $U_0-E<t$, the transmission through a sharp junction is no longer equal to unity at normal incidence, which can be explained by intervalley scattering. When we consider a smooth potential, intervalley scattering is strongly reduced, and we have observed that there is almost total transmission. In the regime where both $E>t$ and $U_0-E>t$, we have found that there is total reflection for both a sharp and a smooth junction. This can be theoretically explained by the different spinor structure of the wave functions in the electron and hole regions.

We have also modeled the scattering of a wave packet by a two-dimensional Gaussian potential. For both a potential barrier and a potential valley a quantum interference pattern is formed. We have compared this pattern with the classical electron trajectories and the associated caustic, and find that the agreement improves when the width of the potential increases.

The numerical scheme developed in this paper is powerful in dealing with large-scale systems. Since the scheme uses the tight-binding model, one has full control over the sample structure and the electronic potential at each atomic site. This enables the study of different types of potential barriers, either single barriers or multiple in an array. Using the TBPM, we can also study scattering due to the presence of disorder like vacancies, adatoms, ad-molecules, charge impurities, local reconstruction (e.g., pentagon-heptagon rings), grain boundaries and local strain or compression. We leave these problems for future work.

\section{Acknowledgments}

We are grateful to Erik van Loon for helpful discussions.
We acknowledge financial support from the European Union Seventh Framework Programme under Grant No. 604391 Graphene Flagship, ERC Advanced Grant No. 338957 FEMTO/NANO, and the Netherlands National Computing Facilities foundation (NCF).

\appendix

\section{Vanishing of higher order terms in perturbation theory} \label{app:Tmatrix}

In this appendix, we will show that scattering between the different eigenstates of the Hamiltonian~(\ref{eq:H_kx_normal}) is forbidden for any scalar potential $U(x)$. To this end we introduce the $T$-matrix (see e.g. Ref.~\onlinecite{Newton86}), which is defined by
\begin{equation}
  \hat{T}=\hat{U}+\hat{U}\hat{G}_0\hat{T}, \label{eq:T-matrix}
\end{equation}
where $\hat{U}$ is the operator of potential scattering, and $\hat{G}_0$ is the free particle Green function,
\begin{equation}
  \hat{G}_0=\lim_{\epsilon\to +0} \frac{1}{E-\hat{H}_0+i \epsilon} . \label{eq:Green0}
\end{equation}
The probability of scattering between the states $|\mathbf{k}\rangle$ and $|\mathbf{k}'\rangle$ is then given by $T(\mathbf{k}',\mathbf{k})=\langle \mathbf{k}' | \hat{T} | \mathbf{k} \rangle$. We can solve Eq.~(\ref{eq:T-matrix}) iteratively, which gives the scattering probability as
\begin{align}
  T(\mathbf{k}',\mathbf{k})&= \langle \mathbf{k}' | \hat{U} + \hat{U} \hat{G}_0 \hat{U} + \hat{U} \hat{G}_0 \hat{U} \hat{G}_0 \hat{U} + \ldots | \mathbf{k} \rangle , \nonumber \\
  &= T^{(1)}+T^{(2)}+T^{(3)}+\ldots
  \label{eq:TMatrixExpFull}
\end{align}
The first term of Eq.~(\ref{eq:TMatrixExpFull}) is just the matrix element in the first order Born approximation that we have seen before in Eq.~(\ref{eq:TMatrixExp}). The other terms are higher order corrections in perturbation theory.

Let us consider scattering of a normally incident electron with an energy outside of the Dirac regime, which is described by the Hamiltonian~(\ref{eq:H_kx_normal}) in momentum space. We will show that for this system scattering between eigenstates with a different spinor structure is forbidden, i.e. that all higher order terms in Eq.~(\ref{eq:TMatrixExpFull}) vanish. The derivation is in the spirit of that in Ref.~\onlinecite{Ando98}. To prove that all terms of the $T$-matrix vanish, let us start by considering $T^{(2)}$. A short calculation shows that it is proportional to
\begin{equation}
  T^{(2)} \propto \int \mathrm{d} q_x \; \chi_{k_x'}^\dagger \, U_{k_x'-q_x} G_{0,q_x} U_{q_x-k_x} \, \chi_{k_x} , \label{eq:T2}
\end{equation}
where $\chi_{k_x}$ denotes the spinor structure of the state with momentum $k_x$. Using the Hamiltonian~(\ref{eq:H_kx_normal}), we find that the free particle Green function in momentum space equals
\begin{equation}
  G_0(q_x)=\frac{1}{E-t f(q_x)\sigma_x+i \epsilon}=
  \frac{1}{t}\frac{|f(q_x)|+ f(q_x) \sigma_x}{(|f(q_x)|+i \tilde\epsilon)^2-f(q_x)^2}. \label{eq:G0ac}
\end{equation}
Since this expression only contains the Pauli matrix $\sigma_x$, we note that Green functions with different arguments commute, and that they have a common eigenbasis. Furthermore, the Fourier components $U_{k_x'-q_x}$ of the potential are proportional to the unit matrix in pseudospin space. Therefore, multiplying the different terms in Eq.~(\ref{eq:T2}), we find that $T^{(2)}$ has the following structure:
\begin{equation}
  T^{(2)} \propto \int \mathrm{d} q_x \; \chi_{k_x'}^\dagger \, (T^{(2)}_0 \mathbbm{1} + T^{(2)}_x \sigma_x) \, \chi_{k_x} , \label{eq:T2v2}
\end{equation}
where $\mathbbm{1}$ is the unit matrix, and $T^{(2)}_0$ and $T^{(2)}_x$ are scalar quantities that depend on the Fourier components $U_{q_x}$ and on the function $f(q_x)$. Now let us consider the situation that $\chi_{k_x}$ is proportional to $(1,1)^T$ and $\chi_{k_x'}$ is proportional to $({-1},1)^T$. Since these vectors are orthogonal, and since they are both eigenvectors of $\sigma_x$ (with different eigenvalues), we see that for this case Eq.~(\ref{eq:T2v2}) vanishes.

In the same way, one can show that all higher order terms in Eq.~(\ref{eq:TMatrixExpFull}) vanish. Since the Green functions for different momenta commute (in pseudospin space), they have a common eigenbasis that consists of the vectors $(1,1)^T$ and $({-1},1)^T$. Therefore, the product of potentials and Green functions also has the structure~(\ref{eq:T2v2}) for higher order terms, and the entire argument runs analogously. Therefore, we conclude that for scattering between a state with spinor structure $(1,1)^T$ and one with $({-1},1)^T$ the $T$-matrix~(\ref{eq:TMatrixExpFull}) vanishes to all orders in perturbation theory. Hence, scattering between such states is forbidden.

\bibliographystyle{apsrev}
\bibliography{BibliogrGrafene}

\end{document}